\documentclass[twocolumn]{openjournal}

\usepackage[utf8]{inputenc}
\usepackage[T1]{fontenc}
\usepackage{float}
\usepackage{ulem}
\usepackage{graphicx}	
\usepackage{amsmath}	
\usepackage{comment}
\usepackage[dvipsnames,svgnames]{xcolor}
\usepackage{color}
\usepackage{hyperref}
\hypersetup{colorlinks=true,
	urlcolor=blue,  
	linkcolor=blue,  
	citecolor=blue, 
    menucolor=blue, 
    urlcolor=blue}  

\def\mpc{\, {\rm Mpc}}

\def\msun{\, M_{\odot}}

\def\simlt{\lower.5ex\hbox{$\; \buildrel < \over \sim \;$}}
\def\simgt{\lower.5ex\hbox{$\; \buildrel > \over \sim \;$}}

\def\redgal{no-break LRD}
\def\redgals{no-break LRDs}
\def\sersic{S{\'e}rsic}
\def\contourtext{The contours contain $1\sigma$ and $2\sigma$ of galaxies in their subpopulation.}
\newcommand{\astrid}{\texttt{Astrid}}
\def\LRDnum{17}
\def\LRDnumwords{seventeen}

\newcommand{\rev}[1]{\textcolor{Black}{#1}}

\begin{document}

\title{The Properties of Little Red Dot Galaxies in the ASTRID Simulation}

\author{Patrick LaChance$^{1,*}$}
\author{Rupert A.C. Croft$^{1}$}
\author{Tiziana Di Matteo$^{1}$}
\author{Yihao Zhou$^{1}$}
\author{Fabio Pacucci$^{2}$}
\author{Yueying Ni$^{3}$}
\author{Nianyi Chen$^{4}$}
\author{Simeon Bird$^{5}$}

\thanks{$^*$E-mail:plachance@cmu.edu}
\affiliation{$^{1}$ McWilliams Center for Cosmology and Astrophysics, Department of Physics, Carnegie Mellon University, Pittsburgh, PA 15213 USA \\}
\affiliation{$^{2}$ Center for Astrophysics $\vert$ Harvard \& Smithsonian, 60 Garden St, Cambridge, MA 02138, USA\\}
\affiliation{$^{3}$ Michigan Institute for Data and AI in Society,
500 Church Street, Suite 600, Ann Arbor, MI 48109\\}
\affiliation{$^{4}$ Institute for Advanced Study, 1 Einstein Dr, Princeton, NJ 08540\\}
\affiliation{$^{5}$ Department of Physics \& Astronomy, University of California, Riverside, 900 University Ave., Riverside, CA 92521, USA\\}

\begin{abstract}

We present simulated counterparts of the ``Little Red Dot'' (LRD) galaxies observed with JWST, using the large cosmological hydrodynamic simulation, ASTRID.
We create mock observations of the galaxies ($5 \leq z \leq 8$) in ASTRID, and find \LRDnumwords\ which fit the color and size criteria of LRDs.
These LRDs are galaxies with high stellar masses ($\rm log(M_*/\msun) \geq 9.7$), and massive black holes ($\rm log(M_{BH}/\msun) \geq 6.8$).
The host galaxies are dense, with stellar half mass radii ($\rm 325\,pc \leq r_{{\rm half},*} \leq 620\,pc$), and dust attenuation in the F444W band above 1.25. Their star formation has been recently quenched.
They host relatively bright AGN that are dust-obscured and contribute significantly to the rest-frame optical red slope and have relatively low luminosity in the rest-frame ultraviolet, where the host galaxy's stars are more dominant.
These LRDs are in an evolutionary phase of miniquenching that is the result of AGN feedback from their massive black holes.
The LRDs in ASTRID are bright with F444W magnitudes of $23.5-25.5$. The less massive and fainter galaxies in ASTRID lack the dust concentration necessary to produce the red slope of an LRD, though this could be an effect of limited resolution.
Most of the highest Eddington black holes are not LRDs due to \rev{ insufficient dust attenuation from their host galaxies, which are also experiencing relatively high star formation rates. This results in their spectra being too flat, despite their highly accreting black holes.}

\end{abstract}


\maketitle



\section{Introduction}
\label{sec:Intro}

One of the most remarkable and unexpected discoveries from early James Webb Space Telescope (JWST) observations has been spatially small, very red objects at high redshift ($z\geq4$) which have been dubbed ``little red dots'' \citep[LRDs;][]{Matthee_2023, Labbe_2023, Kocevski_2023, Kocevski_2024, Harikane_2023}. These objects have compact or point-like morphologies \citep{Baggen_2023, Guia_2024}, display a red color in the observed wavelength bands between $2.0 \mu m$ and $5.0 \mu m$, and a flat or blue color in the $0.5 \mu m$ to $2.0 \mu m$ wavelength range. LRDs appear relatively common in the early universe ($z \geq 5$) with a number density of $\sim 10^{-5}\, \rm Mpc^{-3} Mag^{-1}$, implying they are not extremely rare or unique objects \citep{Kokorev_2024, Perez_Gonzalez_2024, Labbe_2023, Kocevski_2024, Taylor_2024}. In this work we make mock observations of the galaxies in the cosmological simulation \astrid\ \citep{astrid_BHs, astrid_galaxy_formation, Astrid2024} to identify and analyze the LRD galaxy population in this simulated environment.

In addition to their compact morphology, and photometric colors, many of these LRDs have other notable qualities. These include broad Balmer line emission \citep{Kocevski_2024, Kokorev_2024, matthee_2024,Greene_2024}, non-detection or flat spectra in X-ray and infrared \citep{Akins_2024, Ananna_2024, Yue_2024, Maiolino_2024}, and significant Balmer breaks \citep{Setton_2024, Rusakov_2025}. This combination of features has proven difficult to model, as the standard models for both stellar and AGN emission are unable to reproduce some portion of the spectrum \citep{Durodola_2024}. 

\begin{figure*}
    \centering
    \includegraphics[width=2.0\columnwidth]{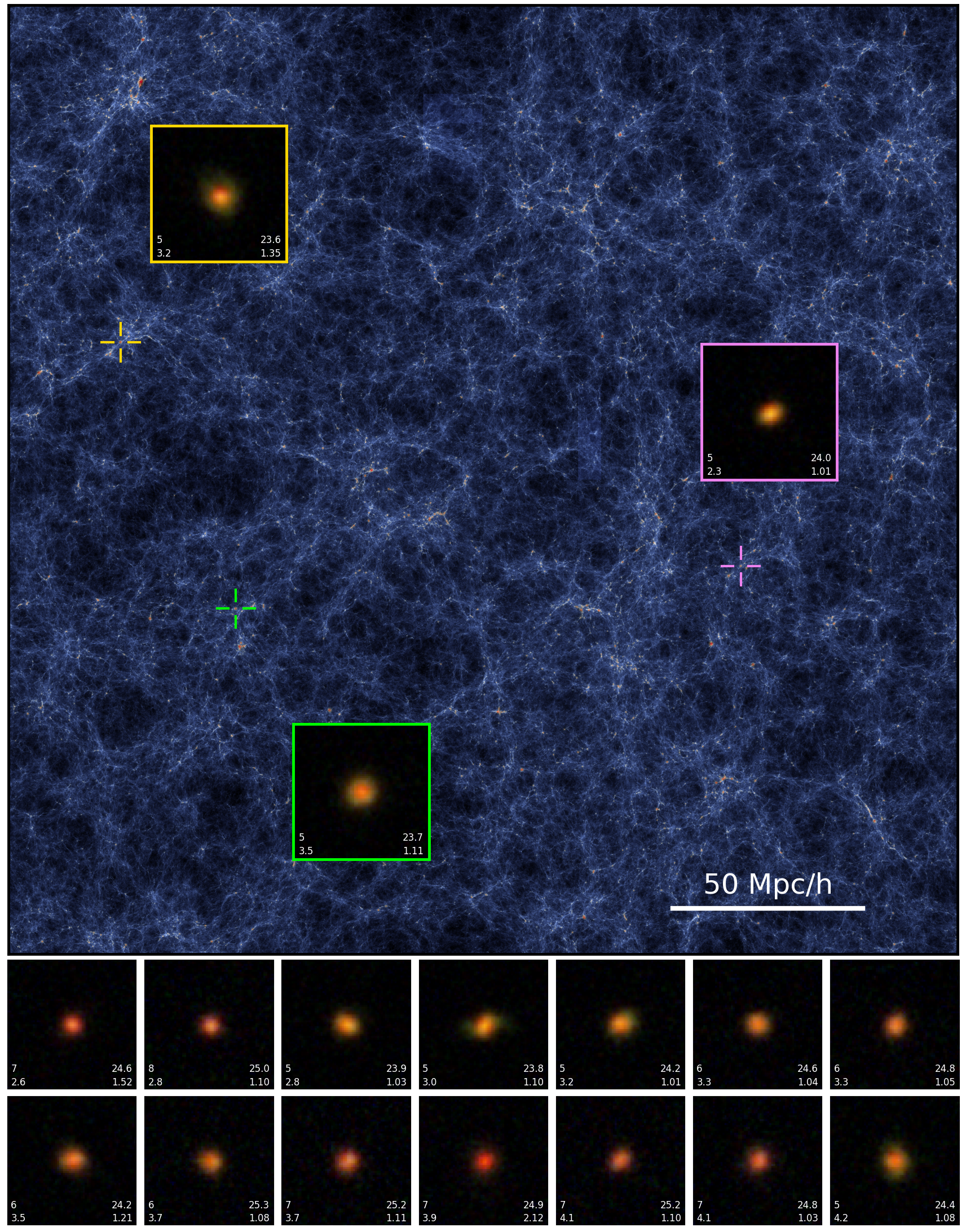}
    \caption{A slice of the \astrid\ simulation at redshift $z=5$, and false color images of the Little Red Dot galaxies in \astrid . The slice is $\rm 25\,Mpc/h$ thick and shows the gas density within the simulation, with color indicating the gas temperature. The three overlayed images are LRDs present in this slice of \astrid\ and their position is indicated by the cross-hair. The false color images use the F444W, F277W, and F150W filter images for the red, green and blue channels, respectively. Each image is 2.0" x 2.0" with pixels of 0.030". The numbers on each image indicate the redshift (top value on the left), $r_{e}$ in pixels (bottom left value), F444W magnitude (top value on the right), F277W-F444W color (bottom right value).}
    \label{fig:galaxy_grid}
\end{figure*}

The broad Balmer emission, combined with their red color, would support an LRD model containing a bright, but strongly dust-obscured, AGN. This model can have issues with the flat emission in the mid-IR wavelength range, and the non-detection of most LRDs in X-ray \citep{Pacucci_Narayan_2024}, as most AGN would have a red mid-IR spectrum, and detectable X-ray emission. Additionally, many attempts to fit these objects as dominated by an AGN project the presence of black holes which are more massive than would be expected for their host galaxy and redshift, or black holes that are accreting well above the Eddington limit \citep{Kocevski_2023, Pacucci_2023, Durodola_2024, Taylor_2024}.

\rev{Recent works in this area have found that models which embed the AGN in a very dense gas cloud can also reproduce some aspects of the LRD spectrum, while avoiding many of the issues associated with other AGN models \citep{Naidu_2025, Taylor_2025, Inayoshi_2025, Jeon_2025}. These ``gas-enshrouded'' AGN models produce the rest-visible red color of LRDs via a very strong Balmer break. This can alleviate the non-detection of LRDs in both the X-ray and IR wavelengths. The dense enshrouding gas which creates the Balmer break, also provides a significant amount of attenuation to any X-ray emission, potentially lowering the resulting X-ray flux below current detection limits. Additionally, the lack of dust attenuation as a source of reddening reduces the amount of dust re-emission that would be expected in the IR bands, which may alleviate tensions with observational detection limits in this region as well.}

It is also possible to model the emission of LRDs as very compact galaxies whose emission is dominated by stellar light. The stellar populations of these models fall into two broad categories: either recent significant starbursts resulting in a younger stellar population in a very dusty galaxy, or earlier starbursts producing an older stellar population with less dust and a significant Balmer break \citep{Perez_Gonzalez_2024, Williams_2024, Baggen_2023, Ma_2024}. These models are generally able to reproduce the visible spectra of LRDs, as well as the X-ray and infrared properties. Depending on the exact stellar population, dust properties, and stellar density used, some models are also able to reproduce the Balmer break, and broad emission line width of observed LRDs \citep{Baggen_2024}.
Early fits of stellar-dominated LRDs produced stellar masses $\gtrsim 10^{10}\msun$ \citep{Akins_2023, Labbe_2023}, placing them in the higher mass ranges for their redshift. Given their point-like appearance, they would have extremely high stellar densities, which could potentially produce the broad line widths that are seen in LRDs \citep{Boylan-Kolchin_2023, Akins_2024, Wang_2024, Endsley_2023, Chworowsky_2024, Baggen_2024, Guia_2024}.

Determining whether the majority of observed LRDs fall into one of these two categories, or somewhere in between, has so far proven difficult, as both the AGN and stellar proposals are able to explain some characteristics of LRDs but not yet all. Additionally, there are many important aspects of our modeling of these galaxies that are uncertain. The two most significant are modeling the AGN spectra and the galactic dust absorption, both of which play very significant roles in determining how we model or fit the observed color and brightness of a galaxy. This has become even more apparent recently as more attention has been paid to the Balmer emission and absorption features of most LRDs \citep{DEugenio_2025, deGraff_2025, Naidu_2025, Inayoshi_2024}, and the dust properties that would be necessary to produce the "V shape" visible spectrum \citep{Setton_2025}. These findings suggest that an emission source with minimal dust obscuration and a very significant Balmer break, such as a gas enshrouded AGN or a compact evolved stellar population, may be the best candidate for the nature of LRDs.

In light of these uncertainties, and the wide range of proposed physical properties (stellar mass, and black hole mass, primarily) for observed LRDs, using simulations to create mock observations of galaxies with known physical quantities and using specific AGN, stellar emission, and dust models may provide some new insight into the nature of LRDs. Having access to the exact stellar and black hole masses that produce specific characteristics allows for new ways to differentiate between the appearance of a star-dominated LRD and an AGN-dominated LRD. Beyond quantities that can be difficult to measure via observation, there is also data available from simulations that is effectively impossible to retrieve from observations. This includes the star formation rate history of a galaxy, and black hole accretion rate history, which could provide evidence for starbursts, or fast black hole growth, which are key aspects of the two proposed explanations for LRD galaxies.

\begin{figure*}
    \centering
    \includegraphics[width=2.0\columnwidth]{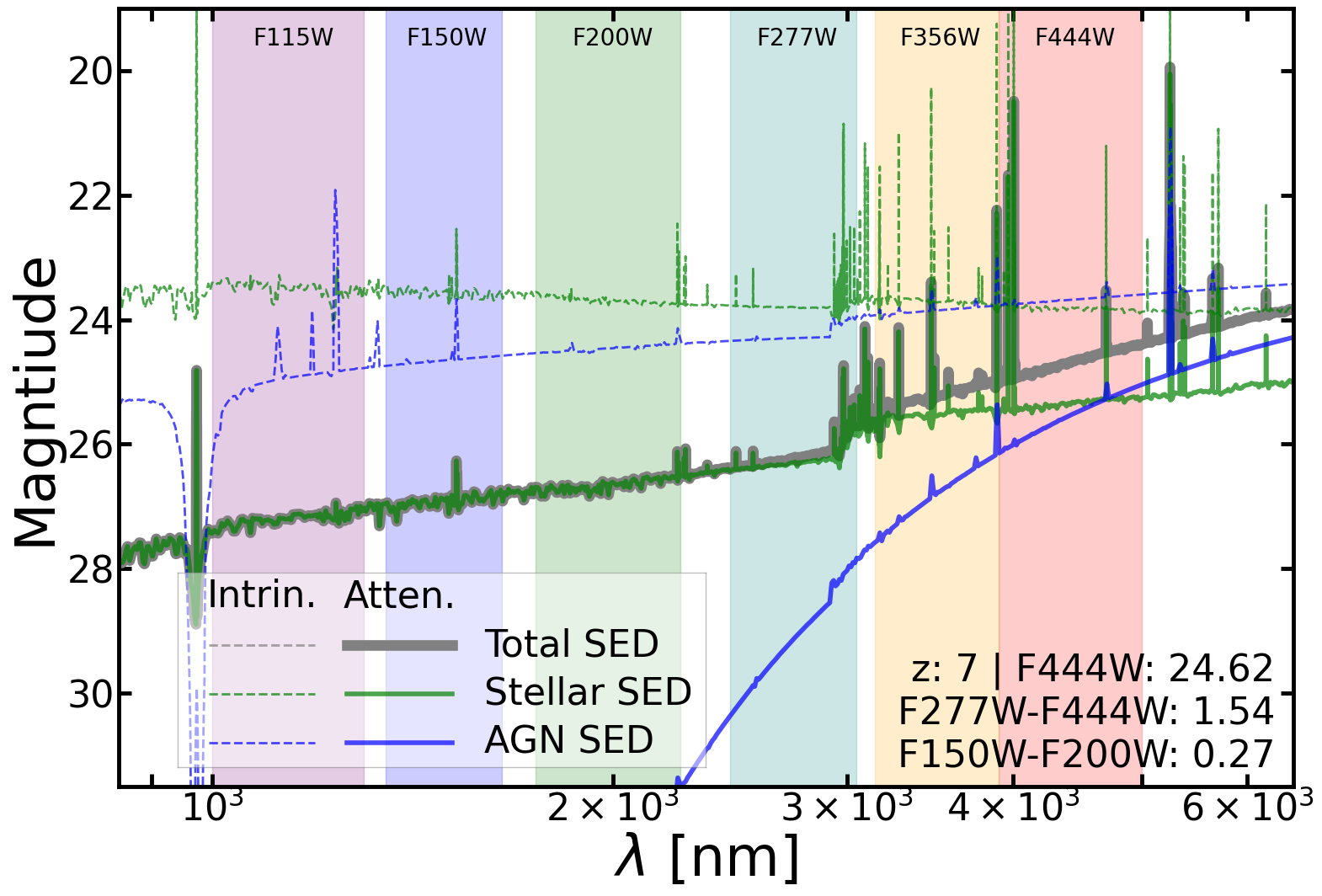}
    \caption{Example spectrum of a little red dot galaxy in \astrid. The solid blue and green lines show the contributions from the AGN and stars respectively, while the thick grey line shows the total observed spectra. The dashed blue and green lines indicate the stellar and AGN spectra before dust attenuation.}
    \label{fig:LRD_spectrum}
\end{figure*}

In this paper, we produce mock observations of galaxies in the \astrid\ simulation between redshifts 5 and 8 (i.e., the bulk of the redshift distribution of the LRDs, as shown in \citet{Kocevski_2024}). We include the emission from both stars and AGN in order to provide a simulation comparison to the current LRD observations. In section \ref{sec:methods}, we describe the process we use to create our mock observations. In section \ref{sec:results}, we describe the population that we produce with this mock observation pipeline, and apply LRD criteria to that population to search for LRDs. In section \ref{sec:analysis}, we examine the properties of the LRDs found in \astrid, how they compare to the overall galaxy population, and possible changes that would produce different LRD populations. Finally, in section \ref{sec:Summary}, we summarize the findings of our work, and discuss future work to enhance our understanding of LRD galaxies.

\section{Methods}
\label{sec:methods}

In this work, we make mock NIRCAM-like observations of galaxies in the \astrid\ simulation in order to identify galaxies that fit the criteria of \rev{LRDs} which are described in detail in section \ref{sec:results}. We build on the imaging process detailed in \citet{Lachance_24}, adding light from AGNs, and the re-emission and dust occlusion of stellar birth clouds.

\subsection{The \astrid\ Simulation}
\label{sec:astrid}

For this work we used the \astrid\ \citep{astrid_BHs, astrid_galaxy_formation, Astrid2024} simulation. \astrid\  has an initial particle load of $2\times 5500^3$ particles in a cosmological box with side-length $250 \mpc h^{-1}$. \rev{This results in a dark matter mass resolution of ($\rm M_{DM} = 9.63 \times 10^6 \msun $) that is similar to the Illustris TNG100($\rm M_{DM} = 7.5 \times 10^6 \msun $) and EAGLE ($\rm M_{DM} = 9.7 \times 10^6 \msun $) simulations}, but with a volume larger than the lower resolution Illustris TNG300 simulation (Illustris TNG: \citealt{Springel_2018, Nelson_2018, Marinacci_2018, Naiman_2018}; EAGLE: \citealt{Schaye_2015}). \astrid\ uses values for the cosmological parameters based on the Planck survey \citep{Planck_2020}. ($\Omega_0 = 0.03089$, $\Omega_{\Lambda} = 0.6911$, $\Omega_b = 0.00486$, $\sigma_8 = 0.82$, $h = 0.6774$, $n_s = 0.9667$). \astrid\ has recently completed its run by reaching  $z=0$.

\astrid\ uses the star formation model described in \citet{Feng_2016}, developed for the \texttt{BlueTides} simulation, and based on the model in \citet{Springel_2003}. Gas particles are half of the initial particle load, and can produce star particles with 1/4 of the initial mass of the gas particle. The criteria for star particle formation, and much of the gas cooling model follows the prescription in \citet{Krumholz_2011}; and \citet{Katz_1996}, respectively. The full details of the handling of gas and star particles within \astrid\ can be found in \citet{astrid_galaxy_formation}. 

The black holes in \astrid\ are seeded within groups of particles identified using the Friends-of-Friends algorithm \citep{Davis_1985}. Once created, they are treated as particles which are subject to gravity, and several forces specific to black holes, including dynamical friction and gas drag \citet{Chen_2022}. This facilitates more physically accurate trajectories and mergers than the more common approach of anchoring the seeded black hole to the galaxy's minimum potential \citep{Chen_2022b}. Black hole accretion is calculated with a Bondi-Hoyle-Lyttleton style model \citep{DiMatteo_2005, Bondi_1952, Bondi_1944, Hoyle_1939}. Our model allows for accretion to exceed the Eddington accretion rate, with a hard limit at twice the Eddington rate. The thermal feedback that results from AGN accretion is also modeled by imparting energy into the gas particles near the AGN at a rate proportional to the black hole's accretion. The full description of the handling of black holes in the \astrid\ simulation is detailed in \citet{astrid_BHs}.

In order to identify and analyze galaxies within the simulation, we must identify substructures within the Friends-of-Friends groups. The subhalo identification algorithm \texttt{SUBFIND} \citep{Springel_2001} is then run to identify galaxy-like substructures in these friends-of-friends groups. The properties and population statistics of these subhalos can be found in \citet{astrid_galaxy_formation}, which showed that subhalo statistics are in good agreement with observational data.

For this work, we focused on the snapshots at redshifts z=5, 6, 7, and 8 as that covers the majority of the redshift range where LRDs have been seen in observations \citep{Kocevski_2024}. We analyze all subhalos in these snapshots with a stellar mass above $10^8 \msun$ as this is near the lowest stellar mass fit from observations. Unlike \citet{Lachance_24}, we do not perform any cuts for observation quality, as we are not analyzing the morphologies of the entire population, and are instead focused on the photometry of the galaxies. 

\subsection{Galaxy imaging process}
\label{sec:imaging_process}

We use the imaging pipeline described in \citet{Lachance_24} as the starting point for our imaging process in order to produce mock observations like those seen in figure \ref{fig:galaxy_grid}. This pipeline uses the star and gas particles present in subhalos found using \texttt{SUBFIND} to calculate the intrinsic stellar emission, and the dust occlusion for each star particle. We use the Binary Population and Spectral Population Synthesis model \citep[BPASS version 2.2.1;][]{Stanway2018} to calculate the intrinsic stellar emission for each star based on its age, mass, and metallicity. The dust occlusion due to the interstellar medium (ISM) for each star at a given wavelength is calculated using the following equation 
\begin{equation} \label{eq1}
\tau_{\rm ISM}(\lambda) = -\kappa_{\rm ISM} \Sigma(x,y,z) \left(\frac{\lambda}{0.55 \mu m}\right)^{\gamma}
\end{equation}
where $\tau_{\rm ISM}(\lambda)$ is the optical depth at a given wavelength, $\kappa_{\rm ISM}$ is a tuning parameter, which we assign a value of $10^{4.1}$
based on the normalization in \citet{astrid_galaxy_formation}, $\Sigma(x,y,z)$ is the metal surface density for that star along the line of sight, and $\gamma$ is a tuning parameter for the slope of the dust model, which we assign a value of $-1.0$ which falls between the slopes of the Small Magellanic Cloud model \citep{Pei_1992} and the `starburst' model \citep{Calzetti_2000}. For further details, see \citet{Wilkins17, Lachance_24}.

We approximate the observed magnitude by summing the luminosities of all star particles that would fall within twice the stellar half-mass radius on the image to approximate the observed magnitude while greatly decreasing computation time.

We analyze the galaxies in wavelength bands that correspond to the JWST filters F444W, F356W, F277W, F200W, F150W, and F115W at their given redshift.

We make two additions to this imaging process in order to emulate emission contributions that are vital to LRDs but were not included in the pipeline used in \citet{Lachance_24}. First, we add the light from galaxy AGNs using spectra produced with \texttt{CLOUDY} \citep{cloudy}. Second, we include the re-emission, and dust occlusion from stellar birth clouds to our stellar dust occlusion model.

\subsubsection{AGN SED calculation}
\label{subsec:AGN_SED}

We use \texttt{CLOUDY} to create our intrinsic AGN SEDs. We use the \texttt{CLOUDY} default AGN input SED, with the default parameters of $-1.4$ for $\alpha_{\rm ox}$, $-0.5$ for $\alpha_{\rm UV}$, and $-1.0$ for $\alpha_X$. We produce 14 different SEDs with temperatures ranging from $10^{4.6}$ K to $10^{6.0}$ K in order to capture the range of temperatures present in the \astrid\ simulation in our redshift range. We produce these SEDs at a set bolometric luminosity of $10^{43.5} \rm erg/s$, and with the following cloudy parameters: inner hydrogen density $= 10^{9}$ $\rm N/cm^{-3}$, inner radius $=  5\times 10^{16} $ cm, total hydrogen column density $= 10^{23}$ $\rm N/cm^{2}$, and metallic abundance of 1/10th the old solar abundance preset. 

We use these AGN SED templates to calculate the SED of individual black holes in the \astrid\ simulation in the following way. We calculate the temperature of each black hole using
\begin{equation} \label{eq2}
T_{\rm BH}(r) = \left(\frac{3 G m_{\rm BH}\dot{m}_{\rm BH}}{8\pi\sigma r^3}\right)^{1/4} (1-\sqrt{r_{isco}/r})^{1/4}
\end{equation}
where $T_{\rm BH}(r)$ is the temperature at distance r from the black hole, $\dot{m}_{\rm BH}$ is the accretion rate of the black hole, and $r_{isco}$ is the radius of the inner most circular orbit for the black hole  ($r_{isco} = \frac{6 G m_{BH}}{c^2}$). Differentiating this equation shows that the maximum temperature occurs at $r = \frac{49}{36} r_{isco}$, and we define the temperature of our black holes to be the temperature at that location. With this temperature, we interpolate between our pre-calculated SEDs to produce the base SED for the black hole. 

\rev{This process models the emission from both the narrow and broad line regions (NLR, BLR) surrounding the AGN, with the inner radius of the gas cloud representing the inner radius of the BLR. Notably, the gas cloud in our model does not include the dynamics of the BLR, which prevents the broadening of emission lines produced in that region. That emission is still included in our model, but is concentrated at the wavelength of the emission line rather than being broadened.}

Because the SED templates are calculated with a bolometric luminosity of $10^{43.5} \rm erg/s$, we also scale the SED up or down in order to match the bolometric luminosity of the black hole present in the galaxy. We calculate each black hole's bolometric luminosity using the equation
\begin{equation} \label{eq3}
L_{\rm bol} = \eta\dot{m}_{BH} c^2
\end{equation}
where $\eta = 0.1$ is the radiative efficiency of the BH.

\rev{We incorporate as much information from the black holes in the simulation as possible when calculating their SED. This includes their mass and accretion rate as described above. Unfortunately, the properties of the surrounding gas cloud are not well defined in the simulation due to their small scale relative to the resolution of the simulation. As a result, we must make some assumptions about these properties. While we choose gas properties that are consistent with standard AGN models, there are recent works which suggest the possibility of significantly denser gas clouds, which would significantly alter the transmitted AGN spectrum \citep{Naidu_2025, Taylor_2025, Inayoshi_2025, Jeon_2025}. We perform an exploratory analysis of a similar model in section \ref{sec:Dim_LRDs}, and conduct an in depth analysis of a similar model in (LaChance et al. In prep.).}

\begin{figure*}
    \centering
    \includegraphics[width=2.0\columnwidth]{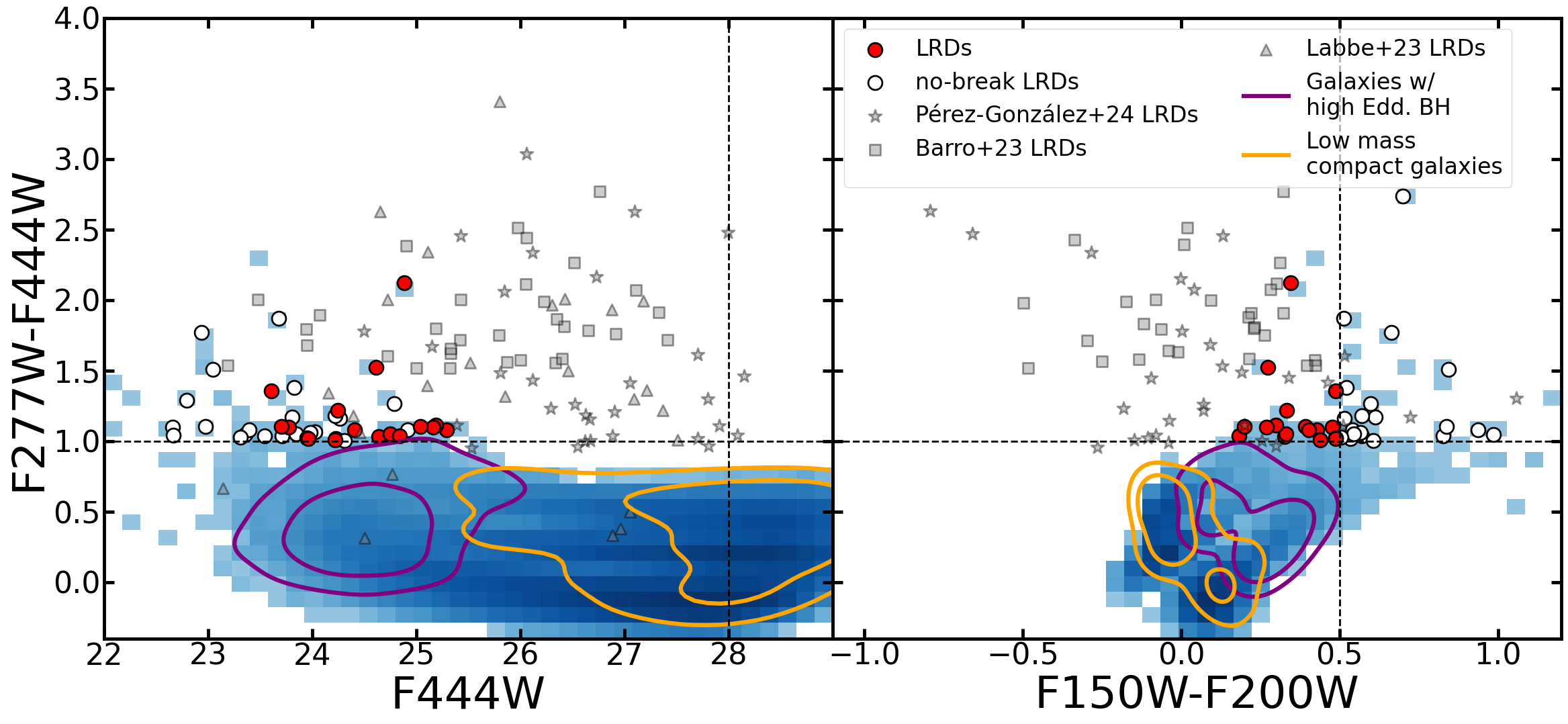}
    \caption{Color-magnitude, and color-color plots showing the main criteria for LRD identification. The red circles indicate galaxies that fit both criteria, while the green squares are galaxies that fit the F277W-F444W criteria, but not the F150W-F200W criteria. \contourtext. \rev{Observational data sets from \citet{Perez_Gonzalez_2024, Barro_2023}; and \citet{Labbe_2023} are included as grey stars, squares, and triangles respectively}}
    \label{fig:LRD_color_criteria}
\end{figure*}

We use equation \ref{eq1} for the dust occlusion of our black holes as well. That said, we change the value of $\gamma$ from $-1.0$ to $-2.0$, producing a dust extinction curve that is more sensitive to wavelength. There is evidence of differences in AGN and stellar dust attenuation laws in LRDs \citep{Brooks_2024, Killi_2024}, and our specific choice of $\gamma = -2.0$ does align with the dust model used in \citet{Ma_2024}. Additionally, we calculate the metal surface density for the black hole on its precise line of sight, using the SPH kernel of each gas particle to calculate its contribution. Specifically, we follow the method described in \citet{astrid_BHs}. We compared this method to the method used for the star particles as described in \citet{Lachance_24} and found minimal difference between the two, with the SPH kernel method being more accurate in cases with very dense, low smoothing length gas particles.

\subsubsection{Stellar birth cloud modeling}

In addition to the dust model for the ISM described in section \ref{sec:imaging_process} \citep{Wilkins17, Lachance_24}, we implement a stellar birth cloud and nebular emission model to our imaging process for this work. We make this addition because one of the leading proposals for star-dominated LRDs is compact starburst galaxies with very young star populations, where emission from birth clouds could be important \citep{Perez_Gonzalez_2024}.

For our nebular emission implementation, we use the built-in nebular emission functionality of \texttt{SynthObs} \citep{wilkins_2020}. Similarly, for the birth cloud occlusion, we follow the prescription from \citet{Vijayan_2021}. This entails calculating the birth cloud optical depth 
\begin{equation} \label{eq_tau_BC}
\tau_{\rm BC}(\lambda) = \kappa_{\rm BC}\left(\frac{Z}{0.01}\right) \left(\frac{\lambda}{0.55 \mu m}\right) ^{-\gamma}
\end{equation}
for star particles that have an age of less than $10^7$ years. $\kappa_{\rm BC}$ is a tuning parameter which fills the same function as $\kappa_{\rm ISM}$ in the ISM model, which we assign a value of $3.0$, and Z is the metallicity of each individual star particle. This optical depth is added to $\tau_{\rm ISM}$ to find the overall optical depth for star particles that fit the age criteria for our birthcloud model.

\subsection{Galaxy Spectra}
\label{subsec:spectra}
In addition to calculating the broadband luminosities that would be observed by JWST, we produce mock spectra of the galaxies in \astrid . These are produced alongside the filter luminosities in \texttt{SynthObs} and are the primary output of \texttt{CLOUDY}, so they have the exact same stellar, AGN, and dust modeling as the filter luminosities we calculate. The spectra for one of the LRDs found in \astrid\ along with a few other types of galaxies are shown in Figure \ref{fig:LRD_spectrum}.

\subsection{Subpopulations of Interest}
\label{subsec:subpops}
In addition to the galaxies which fit the LRD criteria, we also identify two other subpopulations of galaxies within \astrid . First, the population of galaxies which host black holes accreting at or above 50\% the Eddington rate (high Eddington black holes), as they are one of the primary proposed sources of AGN-dominated LRDs \citep{Rusakov_2025, Naidu_2025, Durodola_2024}. Second, the population of galaxies that are compact and low stellar mass. We use the stellar half mass radius ($\rm r_{half,*}$) calculated by \texttt{SUBFIND} as our indicator of compactness, with a criteria of $\rm r_{half,*} \leq 650 pc$. We set the maximum stellar mass to $\rm log(M_*/\msun) = 9.5$. We highlight this population as it should contain the primary candidates for the dimmer LRDs not found in \astrid .

\section{Little Red Dot Identification}
\label{sec:results}

We use this mock observation process to create images of all of the subhalos at redshifts 5, 6, 7 and 8 with a subfind stellar masses above $10^8 \msun$. We use these observations to identify which galaxies fit the criteria of LRDs, and compare them to the LRDs seen in observations. We use criteria comprised of a combination of JWST colors, a minimum magnitude threshold, and a compactness measure, similar to \citet{Kocevski_2024}.

The primary criterion for a galaxy to be selected as an LRD relies on its observed colors. We use the following color and magnitude criteria to identify galaxies as LRD candidates:

$\rm F444W < 28$

$\rm F277W - F444W > 1$

$\rm F150W - F200W < 0.5$

This criteria is based on the color selection in \citet{Perez_Gonzalez_2024}. The magnitude threshold on F444W is necessary to remove any very dim galaxies that may fit the other two criteria due to noise. The $\rm F277W - F444W > 1$ criterion picks out galaxies that are significantly red in the longer wavelength bands of JWST which is necessary to have the red appearance that is core to LRDs. LRDs are not consistently red across the JWST spectrum, but instead have a more neutral color in the shorter wavelength bands. This is indicated by the $\rm F150W - F200W < 0.5$ criteria as it removes galaxies that have a strong red color across both short and long wavelength JWST filters.

While we focus our analysis on the galaxies which fulfill all of the color and magnitude criteria, we also highlight galaxies which fulfill the minimum brightness, and redness criteria (along with the compactness criterion discussed below), but do not fulfill the $\rm F150W - F200W < 0.5$ criteria. These galaxies can provide both a comparison point for the LRD galaxies, potentially identifying the source of the change in slope between the shorter and longer wavelengths, and insight into how important this feature is to LRDs, as some works like \citet{Akins_2024} remove this criterion entirely. We refer to these galaxies as ``\redgals'' as they feature the red appearance of LRDs, but do not have the break in their spectra between the short and long wavelength bands. 

A visual representation of these criteria for our population can be seen in Figure \ref{fig:LRD_color_criteria}. The overall population of \astrid\ galaxies on the $\rm F277W - F444W$ vs F444W axes is largely consistent with that of the overall JWST population on these axes. There is a horizontal ``stripe'' feature, which is the result of our population being produced at integer redshifts, as the Balmer break moves significantly relative to the observed F277W band between redshifts 5 and 6 and again between redshifts 6 and 7. This shift impacts the distribution of $\rm F277W - F444W$ colors between each redshift, resulting in the stripe of density seen in the figure. This is purely an artifact of the integer redshift selection and is not indicative of any discrepancy between our population and the observed population.

In addition to the color criteria for LRDs, we use a size criterion in order to ensure the galaxies have the point-like appearance that is central to LRDs. We take a simplified version of the \sersic\ effective radius criteria \cite[e.g.][]{Kocevski_2024}, which requires that an LRD have a single maximum radius threshold, $\rm r_{e, F444W} \leq 4.5$ pixels. This compactness cut removes 7 of the 24 galaxies which fit the color and magnitude criteria, leaving \LRDnumwords\ total LRDs in the \astrid\ simulation between redshifts 5 and 8. The exact $r_{e}$ value for each LRD can be found in their mock images in figure \ref{fig:galaxy_grid}. When converted to physical units, the median $r_{e}$ is $611$ pc.

The LRDs present in \astrid\ are quite bright, with observed magnitudes in the F444W band of $\rm 23.5-25.5$ ($\rm -20.5 \geq M_{UV} \geq -18.5$), which covers the bright end of the population of observed LRDs. This indicates the LRDs present in \astrid\ likely represent a subpopulation of the observed LRD population which spans an F444W magnitude range of $\rm 22-28$. \rev{\astrid\ also lacks the extremely red objects that are found in observations (F277W-F444W $\geq$ 2.5). These very red objects in observation are generally dimmer ($F444W \gtrsim 25.0$) than the LRDs identified in \astrid. As such, it is unclear if these objects are missing for the same reasons as the other dim LRDs, or if their deficit in \astrid\ is the result of another effect.}

\rev{The low number of LRDs present in \astrid\ relative to observations further supports the fact that we are only capturing a small fraction of the observed LRD population.} Based on the proposed density of LRDs of $\sim 10^{-4}$ $\rm Mpc^{-3}$ from observations \citep{Matthee_2023, Perez_Gonzalez_2024, Kocevski_2024, Kokorev_2024, Finkelstein_2025}, for the volume we are analyzing we would expect to find thousands of LRDs, but we only find \LRDnumwords\ (a density of $\sim 10^{-7}$ $\rm Mpc^{-3}$). \rev{This is due to a combination of the missing subpopulation(s), and an under-density of $\sim$2 orders of magnitude in the bright LRD population (F444W $\leq 25.5 $)} \rev{The mass and luminosity functions of the galaxies and black holes in this redshift range in \astrid\ are largely consistent with observations \citet{astrid_galaxy_formation, astrid_BHs}, indicating this deficit in LRDs is not the result of an under-density of galaxies or black holes in general.} We investigate why more galaxies in \astrid, especially dimmer galaxies, are not LRDs in Section \ref{sec:Dim_LRDs} and discuss further in Section \ref{subsec:Discussion}. 

\rev{The redshift distribution of LRDs in \astrid\ (redshifts 5, 6, 7, and 8 contain 7, 4, 5, and 1 LRDs respectively) is roughly consistent with observations, which find the highest LRD density occurs around $\rm z\sim5-6$ \citep{Kocevski_2024, Pacucci_2025}. Due to the low number of LRDs in \astrid\ it is difficult to make a quantitative comparison with the observational distribution.}

\begin{figure}
    \centering
    \includegraphics[width=\columnwidth]{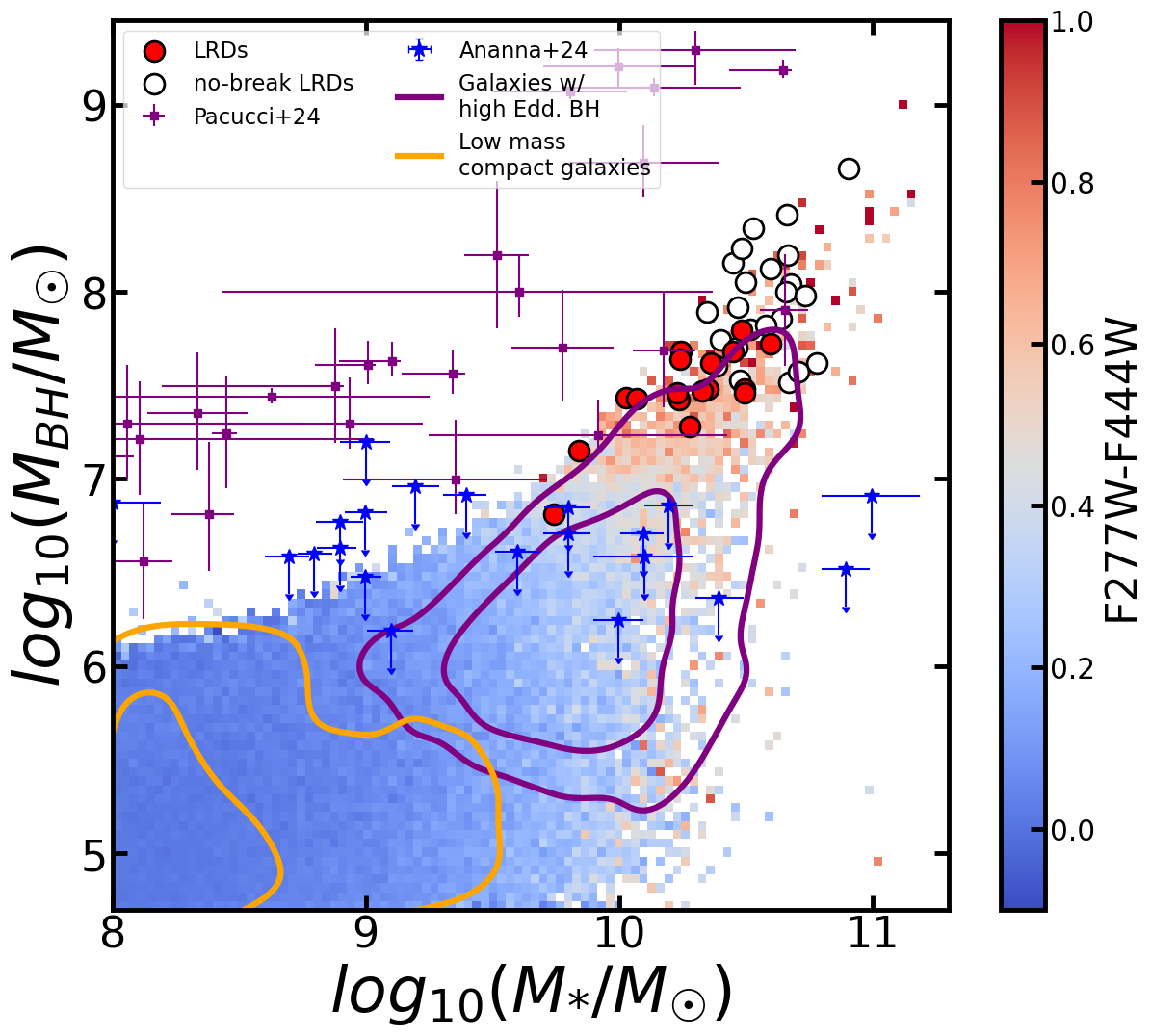}
    \caption{Galaxy black hole mass plotted against galaxy stellar mass. The underlying 2D histogram shows the population of all galaxies in \astrid\ colored by the average F277W-F444W color of galaxies in that pixel, while the red and green points indicate LRDs and \redgals\ found in \astrid\ and in JWST observations. \contourtext\ The purple dataset is comprised of AGN fits from \citet{Maiolino_2024_JADES, Harikane_2023, Ubler_2023, Stone_2024, Furtak_2024, Yue_2024_eiger} compiled in \citet{Pacucci_Loeb_2024}. The blue star dataset was produced in \citet{Ananna_2024} from the X-ray observations of some LRDs.}
    \label{fig:Mbh_Mstar}
\end{figure}

\section{Little Red Dot Properties}
\label{sec:analysis}

We analyze the physical properties of the LRDs present in \astrid\ and compare them with the general galaxy population in \astrid, the two subpopulations of interest, and observations whenever possible, in order to determine the combination of characteristics that result in an LRD.

\begin{figure*}
    \centering
    \includegraphics[width=2.0\columnwidth]{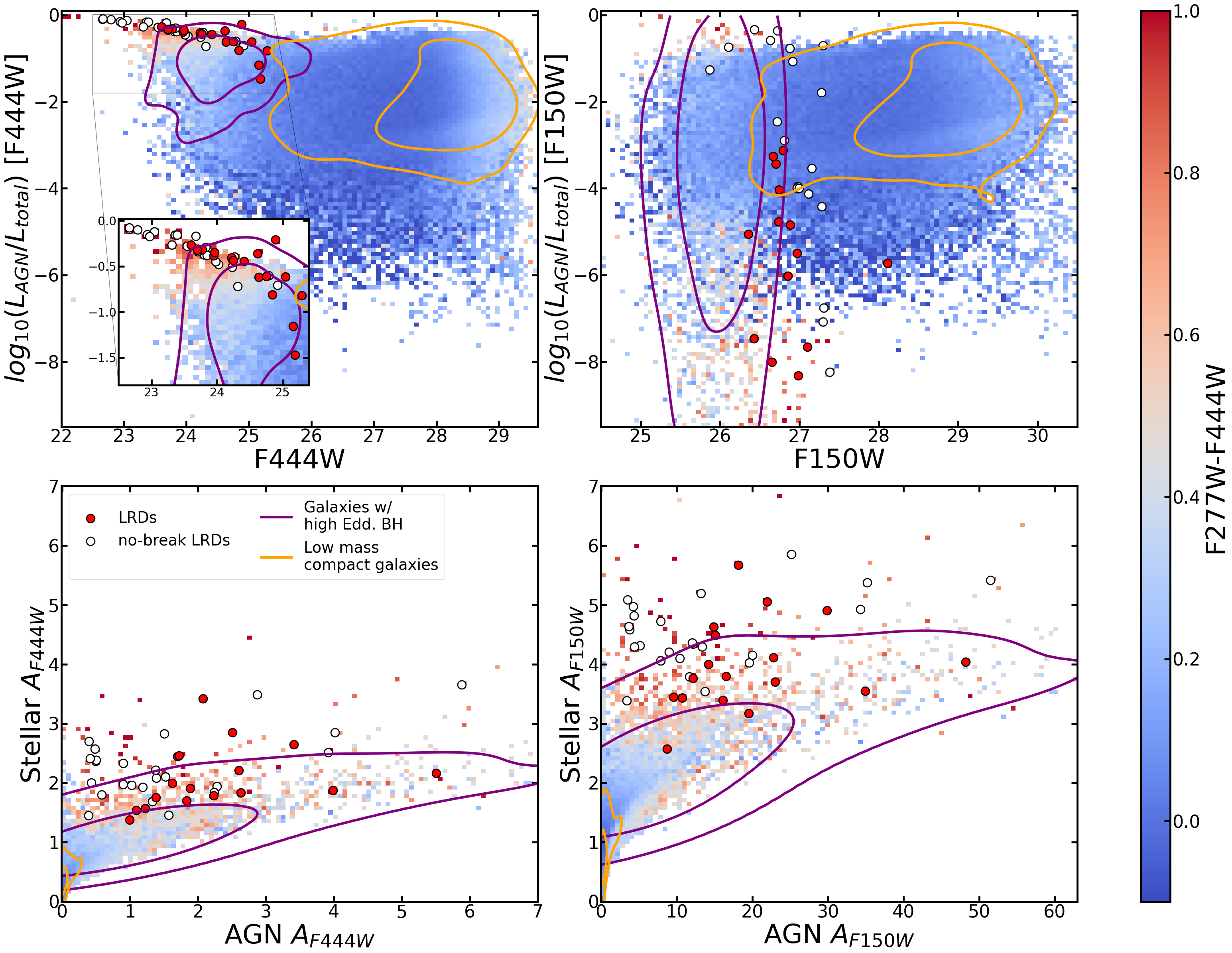}
    \caption{The top two plots show the ratio of AGN to total luminosity vs the magnitude of the galaxy in the F444W band on the left and F150W band on the right. The bottom two plots show the overall attenuation of stellar and AGN light in the same observation bands. The \astrid\ population is presented in the background colored by average F277W-F444W color, while the LRDs are indicated with red circles, and the \redgals\ are shown with green squares. The inset plot on the top left plot zooms in on the region where the LRD and \redgals\ are found. \contourtext}
    \label{fig:L_ratio}
\end{figure*}

\subsection{Galaxy masses}
\label{sec:population_results}

The $\rm M_{BH}-M_{*}$ relation is of particular interest due to recent JWST observations of many black holes which appear over-massive compared to their host galaxies \citep{Pacucci_2023}. We compare our LRD sample with two samples of similar galaxies from JWST observations; first, a sample of galaxies with over-massive black holes, many of which reside in LRDs, aggregated in \citet{Pacucci_Loeb_2024} (over-massive BH sample; shown in purple in Figure \ref{fig:Mbh_Mstar}), and second, a sample from \citet{Ananna_2024} (X-ray sample; shown in blue) which proposes upper bounds on the black hole mass based on the X-ray non-detection of LRDs. While these two samples are not fitting the exact same sources, they illustrate how disparate the proposed stellar and black hole masses for galaxies with similar properties can be, depending on the observational features that are prioritized, and the models used during fitting. This is further supported by the potential for bias in host stellar mass fits found by \citet{Berger_2025}. The dataset compiled in \citet{Pacucci_Loeb_2024} suggests over-massive black holes, most of which cannot be reproduced in \astrid , are present in the LRD population. The dataset from \citet{Ananna_2024} has generally lower black holes masses, and supports star-dominated, or hybrid star+AGN LRDs.

The LRD population in \astrid\ is entirely comprised of galaxies with stellar masses above $10^{9.75} \msun$. This does overlap with the high mass ends of both the over-massive BH sample and the X-ray sample, but completely lacks the lower mass galaxies found in both datasets. 
The LRD galaxies in \astrid\ also contain very massive black holes, on the lower end of the over-massive BH sample, but well above those predicted in the X-ray sample.

Additionally, we see that the high Eddington sources (sources which host black holes accreting at $\geq 50\%$ the Eddington rate) in \astrid\ are present primarily in galaxies with $M_{*} \geq 10^{9.0} \msun$. This is in contrast to many of the models of LRDs, featuring Eddington accreting black holes that are massive relative to their host galaxies, and are present in galaxies of  $10^{8.0} \leq M_{*} \leq 10^{9.0} \msun$.

\subsection{AGN and Stellar Contribution}
\label{subsec:AGN_stellar_cont}


\begin{figure*}
    \centering
    \includegraphics[width=2.0\columnwidth]{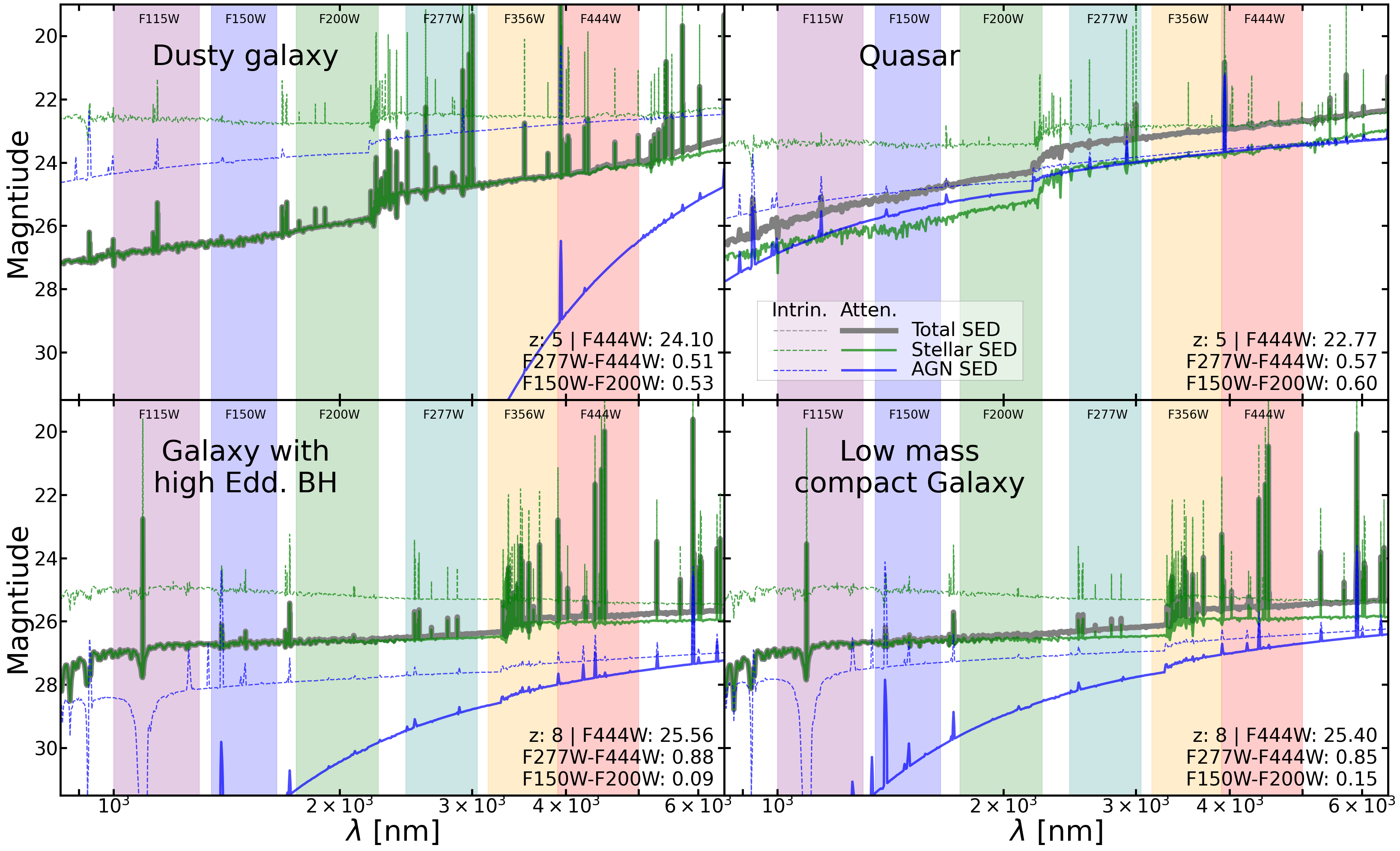}
    \caption{Spectra from four different galaxies which do not meet the LRD criteria. The solid blue and green lines show the contributions from the AGN and stars, respectively, while the thick grey line shows the total observed spectra. The top left panel shows the spectrum of a galaxy with similar stellar and black hole properties as our example LRD in figure \ref{fig:LRD_spectrum}, but which has too much dust attenuation. The top right panel shows a spectrum from another galaxy with similar properties, but with a relatively unattenuated AGN, making it appear as a quasar. The bottom left panel shows the spectrum of a galaxy in our "Galaxies with high Eddington black holes" subpopulation. The bottom right panel shows the spectrum of a galaxy in the "low mass compact galaxy" subpopulation.}
    \label{fig:4_panel_spectrum}
\end{figure*}

\subsubsection{Band-specific contribution and attenuation}

Figure \ref{fig:L_ratio} shows the AGN contribution to galaxy luminosities, and the attenuation of stellar vs AGN light in both long and short wavelength filters. In the F444W band, all but two of the LRDs have an AGN contribution that is at least 10\% of their overall light, with some having AGN that outshine the entire stellar contribution. 
This region of bright galaxies with significant AGN contribution in the F444W band is the only region where the overall population displays a shift towards a redder long wavelength color.

In the F150W band, the majority of LRDs have an AGN contribution below 1\%. There is a population of \redgals\ that feature significant AGN emission in the F150W. These galaxies are similar to the other LRD galaxies, but the line-of-sight to their AGN has significantly less dust relative to the rest of their host galaxy, resulting in it dominating their emission across the JWST spectrum. The dust that is present is enough to redden the AGN spectra, resulting in a consistently red spectra at both short and long wavelengths.

The bottom two panels of Figure \ref{fig:L_ratio} show the attenuation of both the stellar and AGN components of both the overall galaxy population and the LRDs in \astrid. In the F444W band, nearly all LRDs (with the exception of the \redgal\ subpopulation discussed above) have attenuation values ranging from 1 to 4 for both their AGN and stellar components which is consistent with fitted attenuation values from observations \citep{Kokorev_2024, Ma_2024, Akins_2024}. In the F150W band, LRDs have AGN attenuation values of 8 to 60, while the stellar attenuation ranges from 2 to 6. This extreme level of AGN attenuation in the shorter wavelengths aligns with attenuation fits from observation \citep[e.g.][]{Brooks_2024}. This indicates that the increased dust slope relative to wavelength we employ for AGN attenuation is consistent with observations. 

The stellar and AGN attenuation in both the short and long wavelengths for galaxies that host high-Eddington black holes is consistent with the overall galaxy population. The AGN contribution in the F444W of the high-Eddington host galaxies is lower on average than the LRDs. In addition to hosting a high-Eddington black hole, these galaxies must also have other features that facilitate increased AGN contribution.

The population of compact low-mass galaxies includes some AGN, which contribute significantly to their overall spectra, but they feature extremely low dust attenuation. This suggests they require either a non-dust source of long wavelength reddening, such as a very strong Balmer break, or some aspect of the simulation or dust model underestimates the amount of attenuation present in these galaxies.

\subsubsection{Galaxy Spectra}
\label{sec:spectra_results}

We show the spectrum of an LRD galaxy in Figure \ref{fig:LRD_spectrum}, and the spectra of four non-LRD sources in Figure \ref{fig:4_panel_spectrum}. The LRD contains an AGN which contributes significantly to the long wavelength bands, but not to the shorter wavelengths due to dust attenuation. The stellar component is also significantly dust attenuated, and features a noticeable Balmer break which contributes to the overall red slope, while maintaining a flat slope in the shorter wavelengths.

The top two panels of Figure \ref{fig:4_panel_spectrum} show the spectra of galaxies that have similar stellar and black hole masses and luminosities to our LRD population, but fail to appear as LRDs due to their dust content. The top left galaxy has too much dust attenuating both the stars and AGN. This results in a consistent shallow red slope across the spectrum. The top right galaxy is observed along a line-of-sight with minimal attenuation on the AGN. The AGN spectrum is only slightly reddened, and contributes significantly across the JWST spectrum. This indicates how sensitive the appearance of this population of galaxies is to their dust properties. See, in this regard, the model by \citet{Ferrara_2023}.

The bottom two panels show spectra from galaxies in our two subpopulations of interest - galaxies that host high-Eddington black holes, and low mass compact galaxies. The high-Eddington host galaxy is nearly an LRD, due primarily to the Balmer break present in the stellar population. Despite hosting a black hole accreting at 116\% the Eddington rate, the AGN only contributes minimally to the long wavelength spectra due to its unattenuated luminosity being low relative to its host galaxy, and the minimal dust attenuation of its stellar population. The compact low mass galaxy has very similar features to the high-Eddington host galaxy, with a slightly more massive black hole accreting at 58\% the Eddington rate, and a very similar stellar spectrum. This is one of the reddest galaxies that meet the compact low mass criteria. As these galaxies are the prime candidates for the dimmer observed LRDs we are unable to find in \astrid\ we analyze their properties further in section \ref{sec:Dim_LRDs}.

\begin{figure}
    \centering
    \includegraphics[width=\columnwidth]{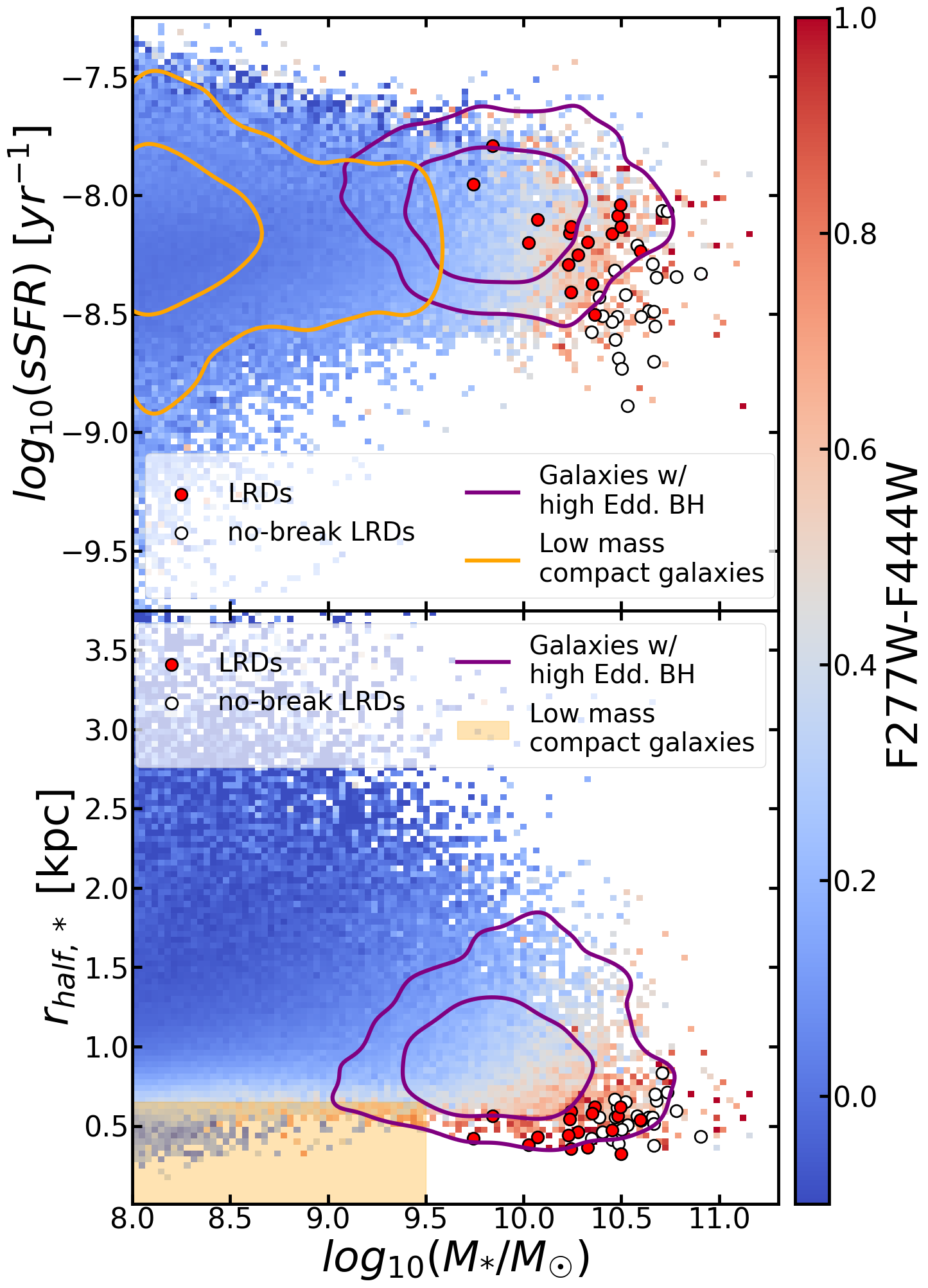}
    \caption{Galaxy specific star formation rate (sSFR) and stellar half mass radius ($\rm r_{half,*}$) vs stellar mass shown in the top and bottom panels respectively. The sSFR is a measure of the star formation rate of a galaxy normalized by its current stellar mass ($\rm sSFR = SFR / M_*$). LRDs present in \astrid\ are marked in red, while \redgals\ are marked in white. The golden shaded region in the bottom panel indicates where the low stellar mass compact galaxy subpopulation resides. \contourtext}
    \label{fig:SFR}
\end{figure}

\begin{figure*}
    \centering
    \includegraphics[width=2.0\columnwidth]{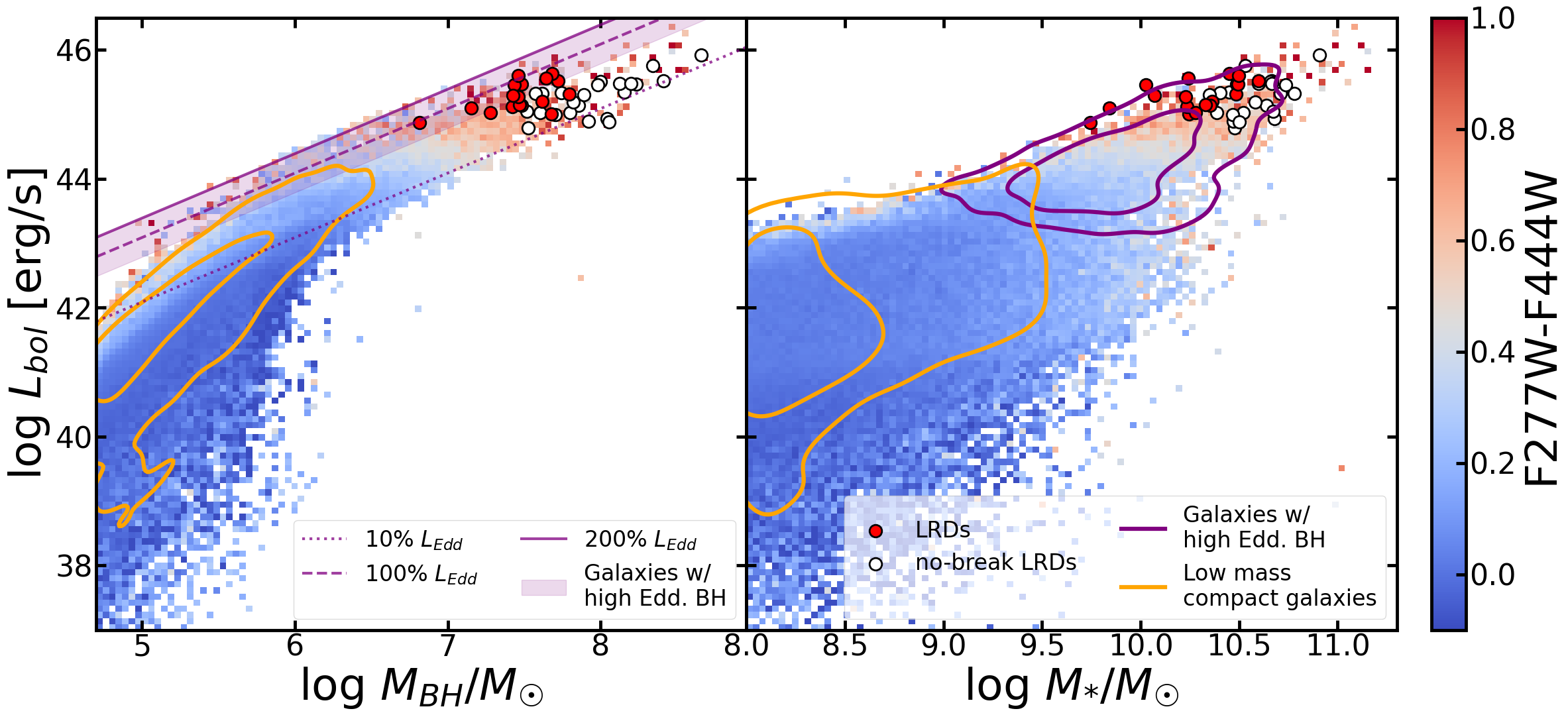}
    \caption{The bolometric luminosity of each black hole plotted against its mass (left panel), and the stellar mass of its host galaxy (right panel). In the left panel, the dotted, dashed, and solid, pink lines indicate 10\%, 100\%, and 200\% of the Eddington luminosity, respectively, and the shaded purple area indicates the area where high Eddington sources are present. LRDs and ``\redgals '' are shown as red and white points, respectively. \contourtext}
    \label{fig:Lbol}
\end{figure*}

\subsection{Stellar and Black Hole Properties}
\label{subsec:stellar_BH_properties}

We examine the properties of the stellar populations and black holes in the general galaxy and LRD populations.

We show the star formation rate (SFR) and stellar half mass radius ($\rm r_{half,*}$) in Figure \ref{fig:SFR}. With the exception of two low mass LRDs, the LRD galaxies have SFRs ranging from typical to well below average for their mass. The bias towards lower SFR among our LRDs suggests they are transitioning towards being quenched likely as a result of feedback from their bright AGN. \citep{Pacucci_Loeb_2024, Pacucci_2024}. Within the LRD population, we see the \redgals\ tend to have lower SFR, which indicates they have been quenched for longer. This is consistent with their redder short-wavelength emission, as their population of young blue stars is smaller.

The two low mass LRDs are observed at redshifts 7 and 8, and are outliers relative to the LRD population in stellar mass and SFR. Additionally, they are among the highest accreting black holes relative to the stellar mass of their host galaxies. This indicates they are still in the phase where cold dense gas is fueling both significant black hole accretion, and star formation, before AGN feedback begins to quench the galaxy. They appear red in the long wavelength bands due to their very bright AGN, and increased dust attenuation from the recent star formation. 

The population of galaxies that host an Eddington accreting black hole tends to have above average star formation for their stellar mass ($\sim 27\%$ more star formation than similar mass galaxies). This indicates the majority of that population is still experiencing star formation, rather than the transition towards quenching that appears in the LRDs. Galaxies in this regime that are also less compact are unlikely to appear as LRDs due to their production of many bright young stars without sufficient dust density to attenuate them.

We confirm the compactness of LRDs via the stellar half mass radius calculated by \texttt{SUBFIND} for each of our galaxies, which is shown in the bottom panel of Figure \ref{fig:SFR}. All of the LRDs present in \astrid\ have compact stellar morphologies, with a median stellar half mass radius of $\rm r_{half, *} = 470$ pc.  \rev{This correlation holds for the overall \astrid\ galaxy population as well, with more compact galaxies having redder long wavelength colors than less compact galaxies of the same stellar mass.} 

We also analyze the properties of the black holes present in the galaxies of \astrid . In the left panel of Figure \ref{fig:Lbol}, we show the bolometric luminosity of the black holes and compare the overall population to the LRDs. Nearly all of the black holes in the mass range where we find LRDs are accreting at a rate of at least $10\%$ Eddington. This is expected as significant accretion over a long period is the main pathway to producing these high mass black holes. The luminosity of the black hole population in this region is consistent with X-ray and UV observational results \citep{astrid_BHs}.

The \astrid\ population overall does show that higher black hole luminosity is correlated with a redder F277W-F444W color. While the luminosity correlation is present across the mass range, the effect is not significant enough to produce many LRDs with lower mass black holes. These lower mass black holes tend to reside in lower mass galaxies, which lack the necessary dust density to sufficiently attenuate both their stellar and AGN emission.

This is supported by the right panel of Figure \ref{fig:Lbol}, which shows black hole bolometric luminosity and galaxy stellar mass. While the same trend towards redder long wavelength colors as black hole bolometric luminosity increases is present across all galaxy stellar masses, no galaxies with stellar masses below $M_* = 10^{9.5} \msun$ are red enough to qualify as an LRD.

\begin{figure}
    \centering
    \includegraphics[width=\columnwidth]{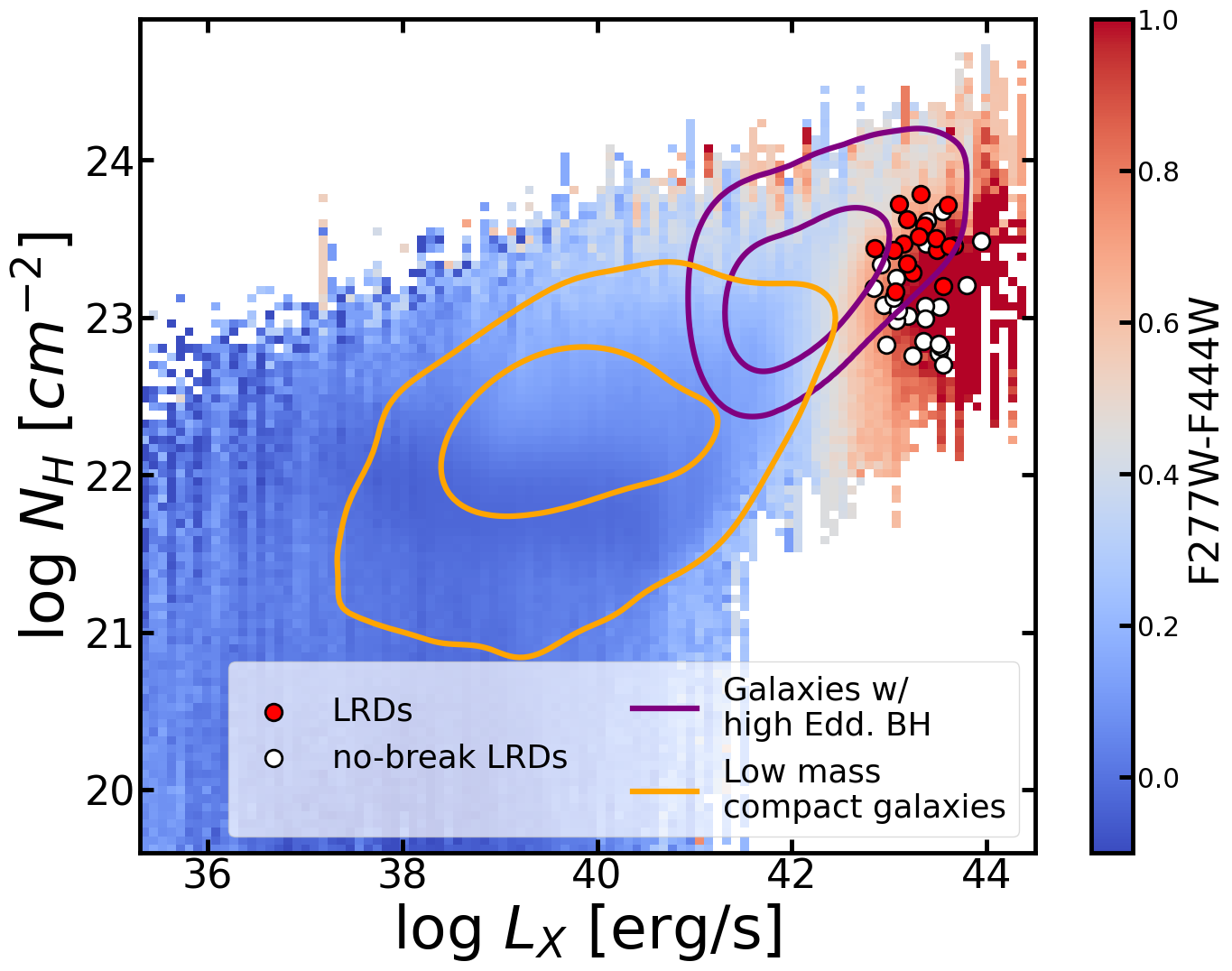}
    \caption{\rev{Column density of ISM Hydrogen along lines-of-sight of the central black hole vs intrinsic X-ray luminosity of that black hole. 50 different lines of sight are calculated for each black hole, all of which are included in this plot. The red and white points indicate the LRDs and \redgals\ respectively, with the NH corresponding to the line-of-sight used for our mock observations. \contourtext\ Notably, the $\rm N(H)$ shown here is solely from the ISM, and does not include any contribution from the local neighborhood of the AGN.}}
    \label{fig:NH}
\end{figure}

\begin{figure*}
    \centering
    \includegraphics[width=2.0\columnwidth]{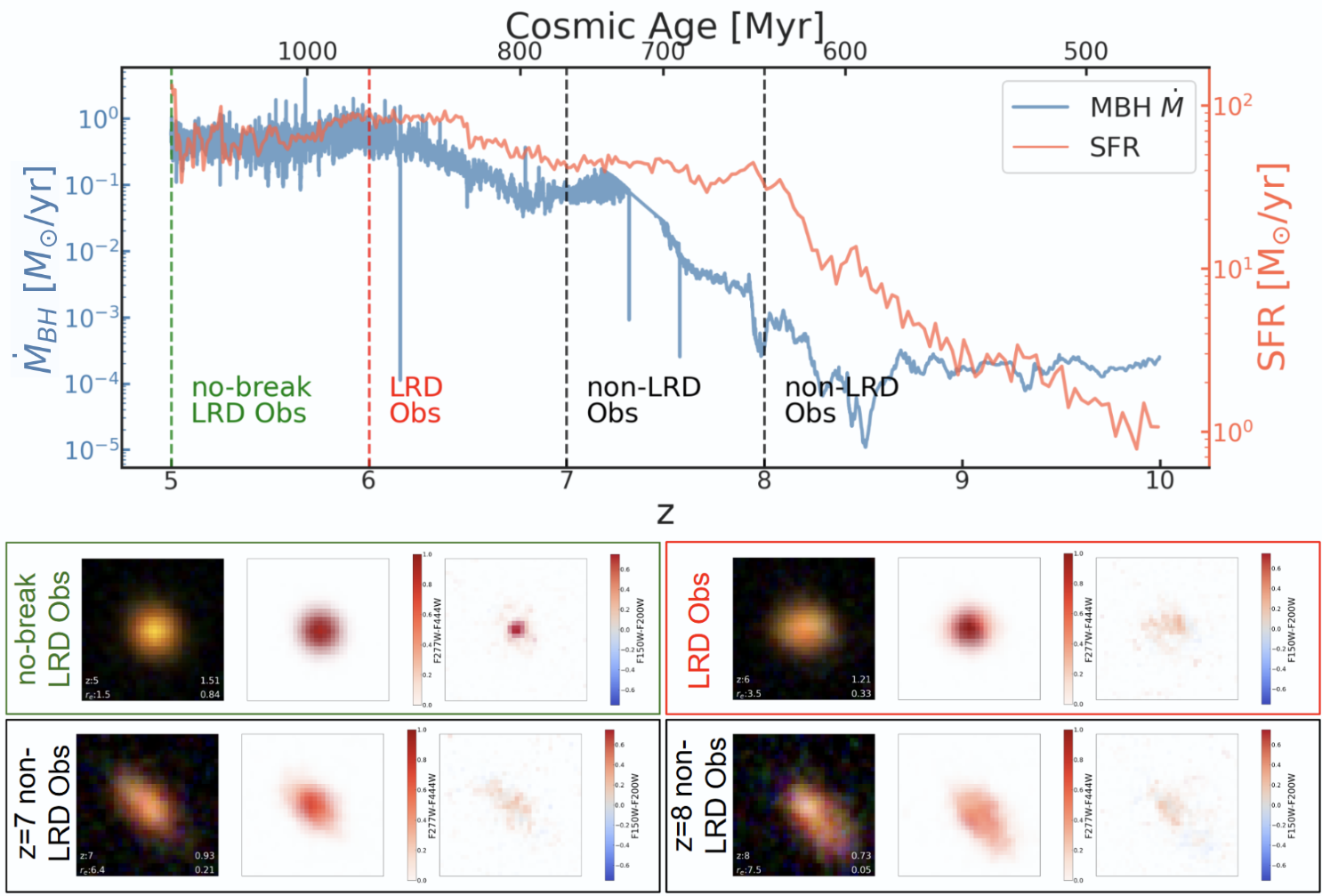}
    \caption{The star formation rate, and black hole accretion rate history of a galaxy in \astrid\ with mock observations made at multiple redshifts. The four mock observations show the galaxy as it appeared at z = 5, 6, 7, and 8 as indicated by the vertical dashed lines. The left values on each false color image show the redshift and \sersic effective radius measured in pixels on the top and bottom, respectively. The right values show the F277W-F444W color and F150W-F200W color on the top and bottom, respectively. The false color images are made from the F444W, F277W, and F150W filters, respectively, and are 1.0" in size. The middle image for each observation is colored based on the F277W-F444W color of each pixel, normalized by the brightness of that pixel in the F444W band. The right image for each observation is similar, but with the color determined by the F150-F200W of each pixel, and normalized by the F200W brightness.}
    \label{fig:SFR_mdot_histories}
\end{figure*}

\subsection{Detectability of the AGN in the X-ray band}

In addition to the observed JWST colors and sizes, one of the hallmark properties of LRDs is that they are generally undetected in the X-rays, even with very deep stacking analysis \citep[see, e.g., ][]{Ananna_2024, Yue_2024, Maiolino_2024, Pacucci_Narayan_2024, Lambrides_2024, Madau_Haardt_2024}. This is, in principle, in tension with the hypothesis that the LRDs contain bright AGN.

Given the heavy obscuration necessary for an AGN spectrum to have the steep red slope in the JWST bands, it is suggested that the lack of X-ray detection may be due to the X-ray emission being dimmed below the detectable range by the same gas and dust that is attenuating the visible light \citep{Maiolino_2024}. \rev{The X-ray emission of the AGN would be attenuated by both the gas in its local environment, and the gas in the ISM along the line-of-sight. The spacial resolution required to fully simulate the AGN environment is well below what is achievable with modern cosmological simulations. As a result we must model this region after the fact via software like \texttt{CLOUDY} as described in section \ref{subsec:AGN_SED}. We calculate the surface density of intervening ISM gas along a line-of-sight from the AGN based on the gas particles of the simulation. This provides a lower bound on the hydrogen surface density that would determine the amount of X-ray attenuation of the AGN, as it does not account for the AGN's local environment. We calculate the hydrogen column density along 50 lines of sight for each black hole in our population, including the line of sight used for our observations, to provide a systematic view of the ISM component of X-ray attenuation of AGN in our LRDs. We show these results in Figure \ref{fig:NH}. We see that the bulk of the LRD  population falls in the $log(N_H) = 23.0-23.5$ range, while the \redgal\ population extends down to $log(N_H) = 22.5$. This is in alignment with the $N_H$ predicted by the Compton thin absorption model proposed in \citet{Akins_2024}, which employs an AGN SED that is weaker in the X-ray region, due to a low $\alpha_{OX}$ ($\alpha_{OX} = -2.0$) \citep{Pacucci_Narayan_2024}. Given the $\rm N(H)$ values we calculate are lower bounds from the ISM only, if we were able to include a self-consistent AGN environment in our calculations, it may raise the $\rm N(H)$ enough such that our results would be consistent with observed X-ray detection limits without relying on an intrinsic X-ray weak AGN spectrum. 
}

Additionally, as discussed in section \ref{subsec:AGN_SED} we adopt a relatively standard X-ray AGN spectrum, with $\alpha_{OX} = -1.4$ and find our LRDs have intrinsic X-ray luminosities in the range $L_X = 10^{43}-10^{44} \rm erg/s$ which falls in the region of the upper limits placed on the X-ray luminosity of LRDs should they be AGN-dominated \citep{Ananna_2024}. Given our population of LRDs is most comparable to the brightest observed LRDs, it is unsurprising that our LRDs have X-ray luminosities near the upper limits from observations. This would also allow for lower mass, dimmer LRDs with similar overall structure to those found in \astrid\ to fall below the upper limits placed on AGN $L_X$ from observational nondetection.

\subsection{When does a galaxy become an LRD: mini-quenching and AGN feedback}
\label{subsec:histories}

Detailed histories of each galaxy's star formation and black hole accretion are available in \astrid\, which we analyze alongside mock observations at different redshifts to track their impact on the galaxy's appearance.

In Figure \ref{fig:SFR_mdot_histories}, we present the star formation and black hole accretion histories of a galaxy which we observe as an LRD at z=6, and again as a \redgal\ at z=5. Beginning at z=8, the galaxy appears slightly red, but not sufficiently to be considered an LRD, and has an extended appearance. Its emission is largely dominated by stellar light at this point, as its black hole is accreting slowly. The red color it does have is due to above-average dust attenuation for its mass.

By z=7, the black hole has a mass of $\rm log(M_{BH})=6.8$ and has begun accreting much more significantly, at 64\% of the Eddington rate. The star formation rate has been roughly constant for 100 Myr. The increase in black hole accretion is not enough to make it contribute significantly to the overall emission of this source, but may have played a role in quenching the starburst that peaked at z=8. The increase in red color from z=8 to z=7 result from a slight increase in the concentration of dust in the galaxy, and the formation of a Balmer break due to the significant population of older stars now present in the galaxy.

Between redshift 7 and 6 the black hole's mass increases to $\rm log(M_{BH})=7.7$ and its accretion rate increases by nearly an order of magnitude (72\% of the Eddington rate). The galaxy underwent another period of increasing star formation followed by a quenched period. These two processes are likely linked via AGN feedback. By redshift 6, the galaxy does appear as an LRD, as it continues to get redder, and appears more compact. The further increase in black hole accretion results in the black hole contributing almost 50\% of the total emission in the F444W band, significantly reddening the appearance of the galaxy. The stellar component maintains a significant Balmer break, and experiences increased dust attenuation due to the galaxy becoming more compact, and continued star formation adding more dust to the interstellar medium. 

Between z=6 and z=5, the black hole accretion remains approximately constant, while the star formation decreases slightly, remaining in the quenched state that began before z=6. This further accentuates the Balmer break and increases the dust density in the galaxy slightly. Additionally, at the time of observation, there is a gap in the interstellar medium along the line-of-sight to the black hole, which significantly increases its brightness relative to its host galaxy. There is still enough dust along that line of sight to redden its emission, resulting in the source having a very red spectrum. This extends to the shorter wavelength bands, as the AGN remains prominent due to its lower relative dust attenuation, shifting this galaxy from being an LRD to being a \redgal .

\section{Where Are The Dimmer LRDs?}
\label{sec:Dim_LRDs}

The population of LRDs in \astrid\ represents a subset of the observed population, as their number density is significantly lower than those found in observations, and they are concentrated within the brighter portion of the observed LRD population. \astrid\ contains $\sim 1/100$ of the expected number of LRDs with UV magnitudes brighter than $\rm M_{UV} = -19$, but completely lacks LRDs with $\rm M_{UV} \geq -18.5$ resulting in an overall LRD number density of $\sim 1/1000$ of observations \citep{Kocevski_2024, Kokorev_2024, Perez_Gonzalez_2024}.

\begin{figure}
    \centering
    \includegraphics[width=1.0\columnwidth]{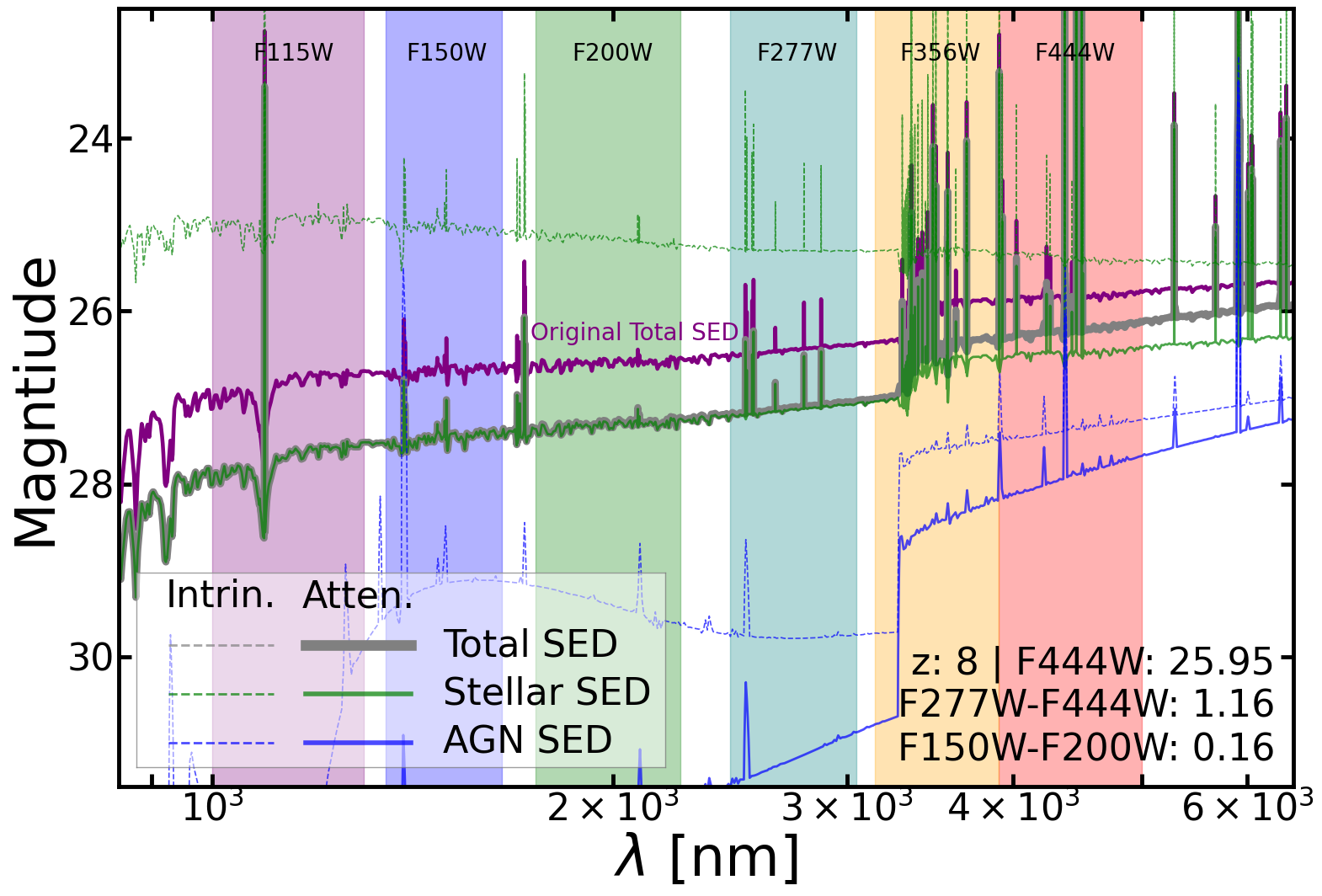}
    \caption{Spectrum of a $M_* = 10^{9.4} \msun$ galaxy hosting a black hole with mass $M_{BH} = 10^{6.1} \msun$, accreting at 116\% the Eddington rate, with the dust occlusion on each star increased by a factor of four and applying an AGN SED model with a Hydrogen column density of $N_H = 10^{24}$ $\rm N/cm^{2}$. This is the same galaxy that appears in the bottom left panel of Figure \ref{fig:4_panel_spectrum}. For comparison we include the ``Total SED'' from that figure (with the original AGN SED and stellar dust models) as a purple line on this plot. With these changes, the galaxy fits the criteria to be an LRD.}
    \label{fig:altered_Edd_LRD_spec}
\end{figure}

We examine the physical and observable properties of the galaxy which produces the spectrum shown in the bottom left panel of Figure \ref{fig:4_panel_spectrum}. It appears dimmer than any of the LRDs within \astrid\ and shares many of the properties of the bright LRDs, which should make it a strong candidate to appear as an LRD. The host galaxy has a stellar mass of $\rm M_* = 10^{9.41} \msun$ and is fairly compact with a half stellar mass radius of $\rm r_{half,*} = 670$ pc. It has a bright black hole, with mass $\rm M_{BH}=10^{6.13} \msun$, accreting just above the Eddington rate, producing a bolometric luminosity of $\rm L_{bol} = 10^{44.29} erg/s$. This results in it having one of the intrinsically brightest AGN of all galaxies with a stellar mass below $\rm M_* = 10^{9.5}$.

This galaxy has an F277W-F444W color of 0.88, which is insufficient to host an LRD. The primary differences between the spectrum of this galaxy and that of the example LRD in Figure \ref{fig:LRD_spectrum} are the relative brightness of the AGN and the dust attenuation of both components. In the LRD spectrum, before dust attenuation, the AGN and stellar components have similar brightness in the F444W band. Despite having one of the brightest AGN relative to its stellar mass among the lower stellar mass galaxies, the AGN emission of the dim LRD candidate galaxy is only $\sim16\%$ as bright as its host galaxy in the F444W band before dust attenuation. This indicates very few, if any, of the lower mass galaxies in \astrid\ have AGN that are bright enough to significantly impact their overall color. This is likely the result of both black hole accretion and star formation being induced by the presence of cold dense gas, producing host galaxies which are brighter due to their increased stellar mass, and young stellar population. This coupling is broken once AGN feedback begins quenching the host galaxy, as we see in our LRDs and models of high redshift galaxies hosting over-massive black holes \citep{Pacucci_Loeb_2024, Pacucci_2024}.

Both the stellar and AGN components of this dim LRD candidate experience significantly less dust attenuation compared to the LRD spectra, which limits the source's red slope. This galaxy is compact and has one of the highest dust attenuation values of the lower mass galaxies, indicating this avenue of spectrum reddening is also limited for our lower mass galaxy population.

These two factors are heavily impacted by modeling choices in our mock observation pipeline. AGN bolometric luminosity is determined by the simulation, but the AGN SED can impact how that luminosity is distributed, and is dependent on the properties of the gas very near the black hole, which is not resolved by cosmological simulations. Recent work \citep[e.g., ][]{Naidu_2025}, suggests that AGN embedded in a denser gas cloud, similar to models of direct collapse black holes \citep[e.g.][]{Pacucci_2015, Pacucci_2016}, may be a strong candidate for LRD emission, which produces a significantly different SED than our more traditional model. Similarly, our dust model has a normalization factor calibrated in \citet{astrid_galaxy_formation}, but may not be tuned appropriately for LRD galaxies.

We experiment with these two modeling choices on our dim LRD candidate galaxy. We increase the hydrogen column density of our AGN SED model in \texttt{cloudy} from $10^{23}$ $\rm N/cm^{2}$ to $10^{24}$ $\rm N/cm^{2}$ to approximate the effect of the ''gas-enshrouded" AGN models proposed in works like \citet{Inayoshi_2025, Naidu_2025} and, \citet{Taylor_2025}. Additionally, we increase the dust normalization by a factor of four (from $\kappa_{\rm ISM} = 10^{4.1}$ to $\kappa_{\rm ISM} = 10^{4.7}$) for the stellar emission to explore the impact of errors in normalization. This also serves as an approximate test of the impact of potential under-density due to simulation effects like resolution limitations and gravitational softening.

The spectrum of the dim LRD candidate galaxy with these changes is shown in Figure \ref{fig:altered_Edd_LRD_spec}. This galaxy does appear as an LRD with these changes, as its $\rm F277W - F444W$ color has shifted from 0.88 to 1.16. With an F444W magnitude of 25.95, and a stellar mass of $\rm M_* = 10^{9.41} \msun$ it would be the dimmest and lowest mass LRD in our population. This indicates it is possible for lower mass galaxies hosting Eddington accreting black holes in simulations to appear as LRDs with specific modeling choices.

\rev{The ``gas-enshrouded" AGN SED introduces another reddening mechanism to these sources, which could also address the lack of very red LRDs in \astrid. Given the ``gas-enshrouded" AGN model we adopted in this experiment was fairly conservative (\citet{Naidu_2025} found a Hydrogen column density of  $\rm N(H)/cm^2 \sim 10^{26}$) it is likely that a more extreme version of this model would have a significantly redder spectrum. If such an SED were applied to Eddington or Super-Eddington sources, they may appear similar to the reddest objects seen in observations. We intend to explore these alternative AGN models in more depth in future work.}

The increased hydrogen column density in the AGN SED introduces both a steeper intrinsic AGN SED and a significant Balmer break, both of which contribute to the red appearance of this galaxy. These qualities are especially important for any LRDs with minimal dust content, like those suggested by \citet{Setton_2025, Xiao_2025}, and would result in the consistent Balmer break transition that is found in \citet{Setton_2024}.

\rev{The increased stellar dust attenuation has two primary effects: (I) it reddens the stellar spectrum and (II) it decreases the brightness of the stellar component, which increases the relative contribution of the AGN to the total emission of the source. This increase in AGN contribution is particularly impactful given the significant red color of the ``gas-enshrouded" AGN SED. If these high Eddington black holes were present in dimmer, lower mass host galaxies, that would also produce sources with heightened AGN contributions.} With the black hole models used in \astrid\ this is unlikely to occur due to the black hole seeding and feedback models, but it may be possible under alternative models. The \texttt{BRAHMA} simulation suite \citep{Bhowmick_2024} suggests that the over-massive black holes seen in observations are reproducible in simulation with specific seeding techniques, and that much of their mass is the result of black hole mergers. This would allow for the formation of more massive black holes in lower mass galaxies, as their mass growth would be less coupled to star formation. These more massive black holes would be capable of brighter AGN emission without exceeding the Eddington limit, potentially appearing as LRDs.

\section{Summary and Discussion}
\label{sec:Summary}

In this work, we look for LRD equivalents in the \astrid\ simulation. We make mock observations of all galaxies with stellar mass above $10^8 \msun$ between z=5 and z=8, which covers the bulk of the redshift distribution of observed LRDs. From this dataset we arrive at the following conclusions:
\begin{enumerate}
    \item Among those observations, we identify \LRDnum\ galaxies that fit all the LRD criteria, and another 24 which have the red color and size of an LRD, but do not meet the flatness criteria in the blue JWST bands (\redgals ).
    \item The LRDs are concentrated in the high stellar mass range, have F444W magnitudes of $\sim 25.5$ or brighter, and UV magnitudes of $M_{UV}=-18.5$ or brighter, representing the bright and massive end of the observed LRD distribution. 
    \item The LRDs have spectra that have significant AGN contribution in the longer wavelength JWST bands, but are dominated by stellar light in the shorter wavelength bands, suggesting they are best fit by a hybrid LRD model.
    \item The LRDs also show suppressed star formation rate compared to the population of galaxies in their mass range, indicating that the feedback from their bright AGNs has begun quenching their host galaxies.
    \item The black holes present in LRDs are massive, with a median black hole mass of $\log(M_{BH}) = 7.5$, accreting at greater than 10\% of the Eddington rate, and heavily obscured.
    \item The primary candidates for the dim LRD population in \astrid\ - low mass compact galaxies, and low mass galaxies hosting Eddington accreting black holes - lack the dust attenuation, and relative AGN brightness of our brighter LRD population.
    \item Adjustments to the dust and black hole models used in the mock observation process, and within cosmological simulations, can produce lower mass and dimmer LRDs.
\end{enumerate}

\begin{figure}
    \centering
    \includegraphics[width=1.0\columnwidth]{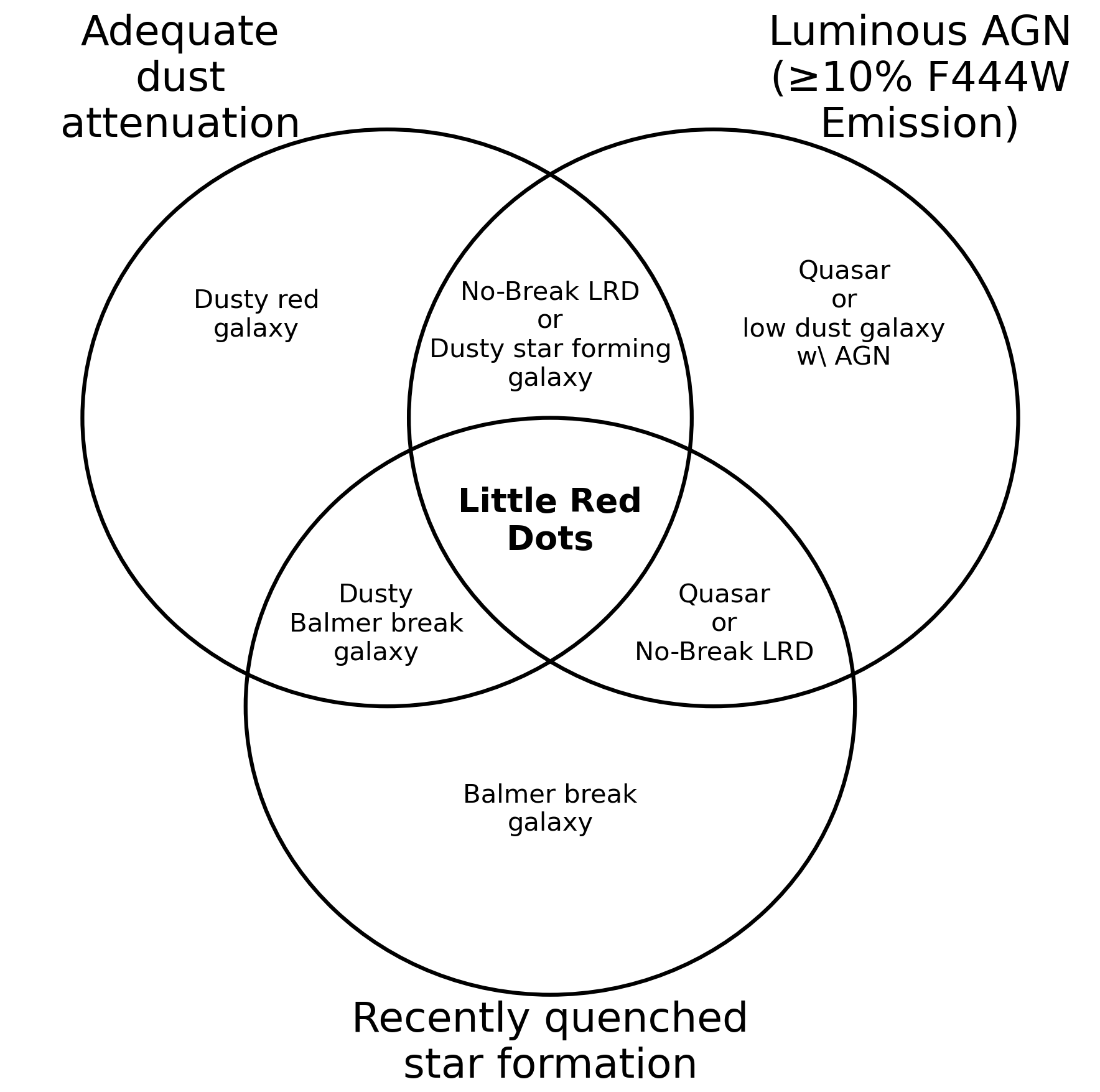}
    \caption{A visual representation of the properties of Little Red Dot sources in \astrid .}
    \label{fig:LRD_venn}
\end{figure}

We outline the properties of Little Red Dots in \astrid\ in Figure \ref{fig:LRD_venn}. There are four LRDs which only satisfy two of these three criteria, two low mass and star forming LRDs discussed in section \ref{subsec:stellar_BH_properties}, and two LRDs with lower AGN contribution discussed in section \ref{subsec:AGN_stellar_cont}.

\subsection{Discussion}
\label{subsec:Discussion}

The size of the \astrid\ simulation is exceptionally advantageous for works such as this one, as it provides a cosmological volume that is much larger than most observational datasets. Given the density of LRDs found by observations across the redshifts we analyze in this work, we would expect to identify tens of thousands of LRDs in our dataset \citep{Kocevski_2024, Kokorev_2024, Matthee_2023, Perez_Gonzalez_2024}. Despite this, we find only \LRDnum\ LRDs, and 24 \redgals\ in our dataset, and those galaxies would be among the brightest LRDs observed.

Due to the compact nature of LRDs, the computational constraints that arise on small scales in cosmological simulations may limit their ability to form. The SPH method used in \astrid\ allows for highly accurate physical modeling on a wide range of scales. Still, it is limited at the smallest scales, due to the mass resolution of particles and gravitational softening. This is relevant for all LRD models, but is of particular importance to the star-dominated models because they are reliant on the existence of extremely dense star-forming regions, which can push the theoretical limits of the current cosmological model \citep{Guia_2024, Ma_2024, Leung_2024, Akins_2024}.

Simulations also must choose specific physical models for processes regarding black holes (e.g., seeding, accretion, and merging) and gas dynamics/radiative properties (e.g., AGN feedback, star formation), etc., all of which impact the possibility of an LRD forming. The choices for these models in \astrid\ are described in detail in \citet{astrid_BHs, astrid_galaxy_formation} and have been shown to have results consistent with observations and other simulations on a wide variety of metrics. AGN-dominated fits of observed LRDs often suggest that over-massive black holes must be present to produce AGNs with sufficient brightness \citep{Pacucci_Loeb_2024, Pacucci_2023, Inayoshi_2024, Durodola_2024}, but most current black hole seeding and accretion models, including those used in \astrid\, do not have the pathways to create such over-massive black holes.

Similar uncertainties arise when analyzing observations of LRDs and trying to determine their properties. Inferred values, like the stellar and black hole masses, are very dependent on the modeling choices made for the stellar population, the AGN SED, and interstellar dust occlusion. This is evident from the spread between the two different observational datasets presented in Figure \ref{fig:Mbh_Mstar}. While these two datasets do not fit the same objects, the extreme differences in black hole mass and stellar mass within the same class of objects indicate a fundamental disagreement arising from the analysis techniques.

Despite refinements in our understanding of the observational properties of these objects, including their dust properties \citep{Brooks_2024, Setton_2025, Xiao_2025, Ferrara_2024, Ferrara_2025_a} and Balmer absorption and emission \citep{Setton_2024, Naidu_2025, DEugenio_2025, deGraff_2025}, there is still a high degree of uncertainty about their nature. This is due to the complex interplay between a variety of their physical properties and the multitude of observational constraints that have been identified for these objects.

Simulations are very well positioned to help alleviate the degeneracy between these properties. Our results are broadly consistent with the observational constraints placed on LRDs. While the LRDs present in \astrid\ do feature significant dust attenuation, they represent the brightest end of the population, indicating the dimmer LRDs may feature less dust attenuation, relying on intrinsic emission that matches the LRD requirements. Additionally, many \astrid\ LRDs feature Balmer breaks from their stellar populations, and the results of section \ref{sec:Dim_LRDs} indicate that dimmer LRDs may feature a higher column density AGN SED that produces a pronounced Balmer break.

Future work in this area should examine the impacts of both the computational constraints and model choices made in simulation. Techniques like zoom-in simulations could be used to create mock observations of potential dim LRDs at different resolutions to determine if their absence in \astrid\ is the result of resolution effects. If such simulations can reduce the range of resolution effects well below the proposed size of star-dominated LRDs (100-200 pc) they would also provide an ideal test bed for evaluating the viability of different models for such LRDs. Similarly, simulations run with different black hole models could provide insight into the conditions necessary for AGN-dominated or hybrid LRD formation. For example, mock observations of galaxies in different simulations within the \texttt{BRAHMA} simulation suite \citep{Bhowmick_2024}, which feature different black hole seeding models, would provide insights into the potential impacts of over-massive black holes on the properties and number density of LRDs. The findings of these works would also help inform the spatial resolution and black hole seeding models used in the next generation of cosmological simulations.

JWST has revealed an entirely new class of astrophysical sources, which show several peculiar properties that are, thus far, puzzling to the AGN and galaxy communities. Cosmological simulations, with their considerable wealth of information for each simulated object, provide a new and exciting pathway to investigate the nature of this puzzling population of distant, compact, and red sources.

\section*{Acknowledgments}
The \astrid\ Simulation was run on the Frontera facility at the Texas Advanced Computing Center.
TDM and RACC acknowledge funding from the NSF AI Institute: Physics of the Future, NSF PHY-2020295, NASA ATP NNX17AK56G, and NASA ATP 80NSSC18K101. TDM acknowledges additional support from  NSF ACI-1614853, NSF AST-1616168, NASA ATP 19-ATP19-0084, and NASA ATP 80NSSC20K0519, and RACC from NSF AST-1909193, SB acknowledges funding supported by NASA-80NSSC22K1897. FP acknowledges support from a Clay Fellowship administered by the Smithsonian Astrophysical Observatory. The Black Hole Initiative at Harvard University also supported this work.


\bibliographystyle{mnras}
\bibliography{references} 

\end{document}